\documentclass[onecolumn,preprint,showpacs,superscriptaddress]{revtex4-1}

\usepackage[utf8]{inputenc}
\usepackage{mathtools}
\usepackage{graphicx}
\usepackage{caption}
\usepackage{subcaption}
\usepackage{xcolor}
\usepackage{soul}
\usepackage{float}
\usepackage{amsmath}
\usepackage{amssymb}
\usepackage{amsfonts}
\usepackage{xcolor}
\usepackage{appendix}
\usepackage{orcidlink}

\begin{document}

\title{Kiselev black strings: the charged rotating solutions}

\author{L. G. Barbosa \orcidlink{0009-0007-3468-3718}}
\email{Corresponding author: leonardo.barbosa@posgrad.ufsc.br}
\affiliation{Departamento de Física, CFM - Universidade Federal de \\ Santa Catarina; C.P. 476, CEP 88.040-900, Florianópolis, SC, Brazil}

\author{V. H. M. Ramos \orcidlink{0000-0002-1763-8663}}
\email{vhmarques@usp.br}
\affiliation{Instituto de Física, Universidade de São Paulo, Rua do Matão 1371,
São Paulo, 05508-090, Brasil}

\author{L. C. N. Santos \orcidlink{0000-0002-6129-1820}}
\email{luis.santos@ufsc.br}
\affiliation{Departamento de Física, CFM - Universidade Federal de \\ Santa Catarina; C.P. 476, CEP 88.040-900, Florianópolis, SC, Brazil}

\author{C. C. Barros Jr. \orcidlink{0000-0003-2662-1844}}
\email{barros.celso@ufsc.br}
\affiliation{Departamento de Física, CFM - Universidade Federal de \\ Santa Catarina; C.P. 476, CEP 88.040-900, Florianópolis, SC, Brazil}

\begin{abstract}
We investigate the properties of a charged rotating black string immersed in a Kiselev anisotropic fluid in anti-de Sitter (AdS) spacetime. The Einstein-Maxwell equations with an anisotropic stress-energy tensor and cosmological constant are analyzed and solved exactly. In this work, we calculate the Kretschmann scalar, obtaining a consistent result that agrees with the existing literature in the absence of charge and fluid. The rotating solution is obtained by applying a coordinate transformation on time and angular coordinates. The event horizon associated with specific values of the equation of state parameter $w_q$ is studied. \textcolor{black}{ The results show an important influence of the fluid parameters  $N_{q}$  and  $w_{q}$ , the charge parameter  $Q$ , and the rotation parameter $a$  on the size of the black string horizon.} \textcolor{black}{In addition, we determine the conditions for the existence of closed timelike curves (CTCs) and compute the conserved charges, such as mass, angular momentum, and electric charge of the black string.} Utilizing the Klein-Gordon equation, we employ the quantum particle tunneling approach to obtain the probability of charged scalar particles tunneling across the event horizon. \textcolor{black}{We obtain the correspondent Hawking temperature as a consequence. Furthermore, we examine the thermodynamic properties, including entropy and heat capacity, to assess the effects of the quintessence field and charge on the black string. The results include particular cases such as the Lemos black string, providing a broader view of black string configurations in AdS spacetime.}
\end{abstract}

\maketitle

\section{Introduction}\label{secI}
Black strings \cite{Lemos:1994xp} are an interesting class of solutions of the Einstein-Maxwell equations, with a negative cosmological constant, considered as counterparts to black holes in AdS spacetimes with cylindrical symmetry. In fact, it was suggested that the model of collapse of a cylindrical system
can be used to mimic a prolate collapse \cite{thorne1972magic}, which may present a physical behavior different from the usual spherical collapse. Such solutions with cylindrical symmetry contrast with hoop conjecture, which states that the collapse of a massive star will produce a black hole only compacted into a region whose circumference in every direction
is $C < 4\pi M$ (we use natural units where $c = G = 1$). However, it was observed \cite{Lemos:1995cm} that this conjecture was formulated considering a spacetime without a cosmological constant. By adding this constant term to the field equations, the differential equations provide cylindrical symmetry
solutions associated with black holes or black strings. In this way, it is expected that the presence of a negative cosmological constant plays an important role in the scenario of cylindrical collapse.   

Shortly after the pioneering work presented in \cite{Lemos:1994xp}, on the solution of black holes with cylindrical symmetry, their charged and rotating version was found \cite{Lemos:1995cm}, and then various configurations were studied, considering different physical and thermodynamic properties \cite{Cunha:2022kep}. Recently, the black string has been investigated in the context of an anisotropic fluid \cite{Ali:2019mxs,Deglmann:2025mcl}, which is a possible candidate to explain the nature of dark energy \cite{perlmutter1999measurements,tsujikawa2013quintessence}. It is worth mentioning that the first solution for black holes in the Kiselev anisotropic fluid background was obtained in \cite{Kiselev:2002dx}, and its charged counterpart was found in \cite{AzregAinou:2014lua}. Furthermore, the charged black string immersed in a Kiselev anisotropic fluid has not yet been determined in the literature. \textcolor{black}{Thus, the Kiselev anisotropic fluid has been studied in a series of works in the literature that aim to understand its effect on the physical systems of interest. One of the main appeals of the solutions obtained in the formulation initially proposed by Kiselev \cite{Kiselev:2002dx} consists in the generalization of several known solutions into a single general solution. On the other hand, some particular solutions describe black holes surrounded by fluids associated with the cosmological constant and quintessence. In a cosmological context, such fluids may be associated with the accelerated behavior of the expansion of the universe due to effects of an equation of state with negative pressure \cite{Kiselev:2002dx}. In this context, it is natural to consider the effects of the anisotropic Kiselev fluid around black strings \cite{Ali:2019mxs} since this class of solution shares many similarities with black holes. Recently, Kiselev's original solution was generalized by considering that the parameter of the equation of state is variable \cite{santos2024regular}. In this way, many solutions known in the literature such as those associated with regular black holes can be reproduced in this context, making the initial formalism proposed in \cite{Kiselev:2002dx} more effective.}

Classically, black holes do not emit radiation; however, this changes if we incorporate quantum effects into the description. In the 1970s, Bekenstein related the properties of black holes to the laws of thermodynamics \cite{bekenstein1973black}, and Hawking showed that under quantum effects, black holes must radiate particles \cite{hawking1975particle,hawking1974black,gibbons1977cosmological}. This discovery provided a new perspective on the quantum theory of gravity \cite{gambini2014hawking,medrano2000hawking}. Subsequently, researchers showed great interest in the field of black hole physics, using different methods to investigate the thermal radiations from these objects \cite{kraus1994simple,srinivasan1999particle,shankaranarayanan2002hawking, ding2009coordinate,Parikh:1999mf,Chakraborty:2015hna,Chakraborty:2015wma}. A particularly interesting method is the quantum tunneling \textcolor{black}{approach} \cite{Parikh:1999mf}. \textcolor{black}{In this manuscript}, using the WKB approximation, it is possible to calculate the tunneling probability from inside to outside the event horizon. Using the Boltzmann factor, we can express the Hawking temperature for an object with a well-defined event horizon, such as black holes \cite{Angheben:2005rm,Christina:2022isq,Wang:2014cza,Rizwan:2016ldf, Kerner:2007rr,Jiang:2005ba, Kerner:2008qv,Javed:2019urt} and black strings \cite{Xu:2016jod,Ahmed:2011qh,Gohar:2011np,Eslamzadeh:2020rbd,Chen:2014xsa,Jusufi:2015mii,Feng:2017lkn,daRocha:2014dla}. \textcolor{black}{Furthermore, in addition to the tunneling of spin-0 particles, particles with different spins have also been studied in this context, including spin-1/2  \cite{Kerner:2007rr}, spin-1  \cite{li2015massive}, spin-3/2  \cite{yale2009gravitinos}, and spin-2 \cite{sakalli2016black}.}

Quantum effects in systems with different backgrounds is a subject that has been studied in several works in the recent years. Particles in cosmic strings backgrounds \cite{Santos:2017eef}, spin-0 fields with the presence of noninertial effects \cite{Santos:2016omw}, rotation effects on scalar fields \cite{Vitoria:2018its}, \textcolor{black}{and quantum effects in magnetic universes, such as the Bonnor-Melvin solution with a cosmological constant \cite{Barbosa:2023gxl,Barbosa:2023rmq}}, are some examples. The important work about Dirac particles near Kerr black holes \cite{chandra} also contributes to this discussion. Other interesting examples may be found in \cite{Sedaghatnia:2019xqb}, that studies Dirac particles in Som-Raychaudhuri spacetime, in \cite{Guvendi:2022uvz}, where the fermion-antifermion system in a spacetime with a topological defect is considered, or the Aharonov-Bohm effect for scalar fields in a spacetime with screw dislocation \cite{Vitoria:2018mun}. So, this kind of formulation may be incorporated in this work.

In this article, we consider the Einstein-Maxwell field equation in the presence of a negative cosmological constant and obtain a solution for a charged and rotating black string immersed in a Kiselev anisotropic fluid in anti-de Sitter (AdS) spacetime. We use the Hamilton-Jacobi ansatz to study the tunneling of scalar particles in this spacetime. In this method, the WKB approximation is employed to solve the wave equation. For this study, we will solve the Klein-Gordon equation for a charged scalar particle, assuming a solution that takes into account the symmetries of the background spacetime. \textcolor{black}{By solving the equation directly}, it is integrated using complex integration around the horizon of the black string.

\textcolor{black}{The structure of this article is as follows: In Section \ref{secII}, we present a new exact solution in this context, generalizing the results of \cite{Ali:2019mxs} by introducing an electromagnetic stress-energy tensor and obtaining an exact solution for a rotating and charged black string surrounded by a quintessential fluid. We analyze the behavior of the solution, compute the Kretschmann scalar, and examine the energy conditions for the solution. Section \ref{secIII} explores the event horizons in the cases of interest, their locations are determined as well as the critical radius for closed timelike curves (CTCs), extreme cases, and the angular deficit. Section \ref{secIV} examines the conserved charges, including the mass, angular momentum, and electric charge of the black string. Section \ref{secV} examines the solution to the Klein-Gordon equation for a charged particle in the spacetime of a charged and rotating black string immersed in an anisotropic fluid, and the Hawking temperature is calculated using the particle tunneling approach. Section \ref{secVI} investigates the thermodynamics of the solution, including the entropy and heat capacity. Finally, in Section \ref{secVII}, we present the conclusions. In Appendix \ref{Riemann_Tensor_Components}, we present the non-vanishing components of the Riemann tensor. In Appendix \ref{Metric_Transformation}, we discuss a transformation that diagonalizes the black string metric. In Appendix \ref{Relativistic_Hamilton_Jacobi_Equation_via_the_WKB_Method}, we derive the relativistic Hamilton–Jacobi equation using the Wentzel–Kramers–Brillouin (WKB) method.}

\section{The equations and the solution}\label{secII}

In this section, we will show how the metric for rotating Kiselev black strings may be obtained and then analyze its main features. The solution for a static charged black string immersed in an anisotropic Kiselev fluid will be determined, and then, by using a transformation of coordinates, the rotation will be implemented. The existence of event horizons and their locations will be studied by performing a variation of the parameters of the theory.

We begin by considering a system under the influence of the Kiselev anisotropic fluid \cite{Kiselev:2002dx,Ali:2019mxs} in addition to the cosmological constant and the Ricci scalar. Besides, we consider the presence of an electromagnetic
field, so, as a result of these assumptions, the action for this system can be written in the form
\begin{equation}
        S = \frac{1}{16\pi}\int{d^4\sqrt{-g}(R - 2\Lambda)} - \frac{1}{16\pi}\int{d^4\sqrt{-g}F^{\mu\nu}F_{\mu\nu}} + S_M,
\end{equation}
\textcolor{black}{where $R$ is the Ricci scalar, $\Lambda$ is the cosmological constant, $S_M$ is the action associated with the fluid}, and $F_{\mu\nu}=\partial_{\mu}A_{\nu}-\partial_{\nu}A_{\mu}$. At this point, we shall focus our attention on the Einstein field equations obtained by varying action with respect to $g_{\mu\nu}$

\begin{equation}
	R_{\mu\nu}-\frac{1}{2}Rg_{\mu\nu}+\Lambda g_{\mu\nu}=8\pi T_{\mu\nu},
\end{equation}
assuming a negative cosmological constant $\Lambda=-\frac{3}{l^{2}}$, where $l$ is the radius of curvature in the AdS spacetime.  We also consider two sources for the energy-momentum tensor, supposing that they do not interact with each other, $T_{\mu\nu}\equiv T_{\mu\nu}^{\left(1\right)}+T_{\mu\nu}^{\left(2\right)}$, with $T_{\mu\nu}^{\left(1\right)}$ given as the electromagnetic energy-momentum tensor
\begin{equation}\label{TensorEMCorda}
	T_{\mu\nu}^{\left(1\right)}=\frac{1}{4\pi}\left(F_{\mu}^{\rho}F_{\nu\rho}-\frac{1}{4}g_{\mu\nu}F_{\alpha\beta}F^{\alpha\beta}\right),
\end{equation}
 with the electromagnetic potential given by $A_{\mu}=-h\left(r\right)\delta_{\mu}^{0}$. The second contribution $T_{\mu\nu}^{\left(2\right)}$ is given by the Kiselev anisotropic fluid energy-momentum tensor \cite{Kiselev:2002dx,Ali:2019mxs} 
\begin{equation}\label{TensorQuiI}
T_{t}^{t}=T_{r}^{r}=-\rho_{q},\qquad T_{\phi}^{\phi}=T_{z}^{z}=\frac{1}{2}\rho_{q}\left(3w_{q}+1\right),
\end{equation}
where $w_q$ is the parameter of the equation of state and $\rho_{q}$ is its energy density. The equation of state given by
\begin{equation}
	p=\frac{1}{2}\rho_{q}\left(3w_{q}+1\right),\qquad -1 \leq w_{q} \leq -\frac{1}{3},
 \label{eqq1}
\end{equation}
is associated with an anisotropic fluid and can be used to describe different types of physical systems, depending on the choice of parameter $w_q$. As observed in \cite{Kiselev:2002dx}, we can mimic well-known solutions in general relativity by choosing particular values for this parameter. For example, the value $w_q = -1$ is associated with a solution with an effective cosmological constant \cite{Kiselev:2002dx}. In the following, we analyze the role of particular values of $w_q$ in this paper.

\begin{figure}[!htb]
     \centering
       \includegraphics[width=0.9\textwidth]{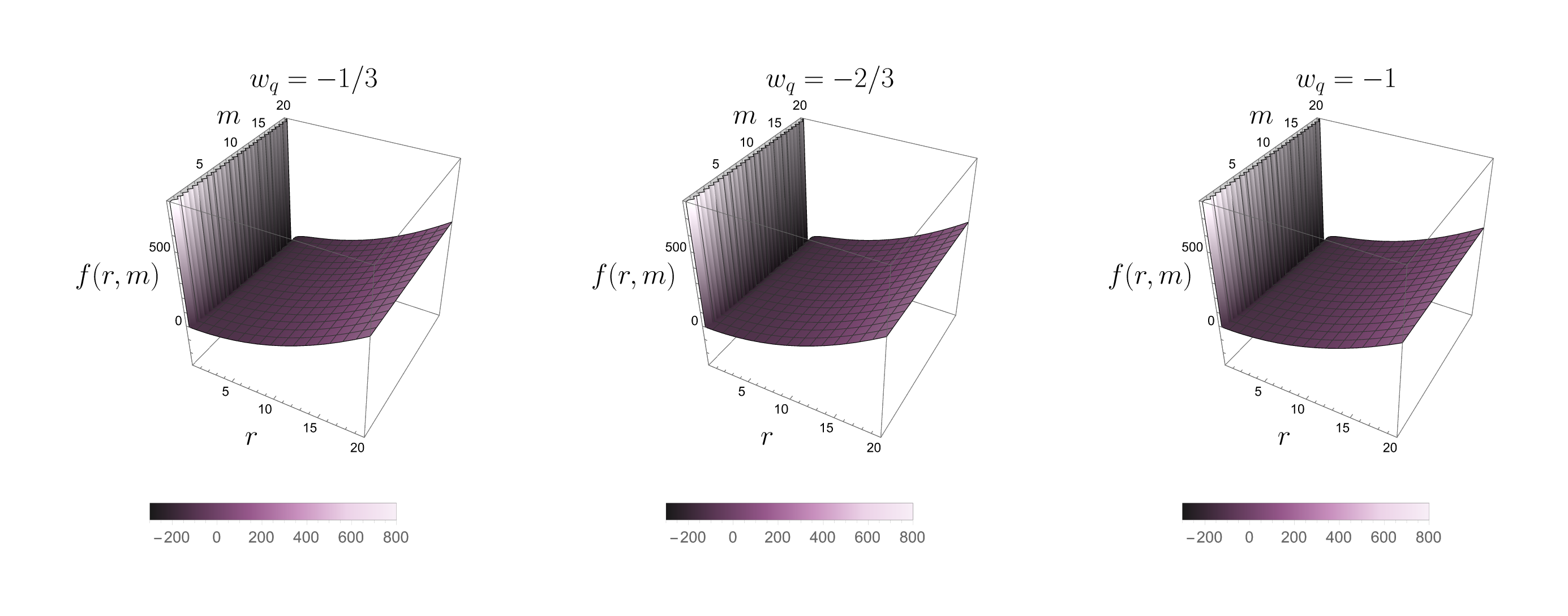}
        \caption{Plot of the event horizon with respect to the radial variable $r$ and the parameter $m$ with $Q=1$ and $N_q=-1$. $N_{q}$ is given in units of $l^{3w_{q}+1}$ and $r$ in units of $l$}
        \label{f1}
\end{figure}

\begin{figure}[!htb]
     \centering
        \includegraphics[width=0.9\textwidth]{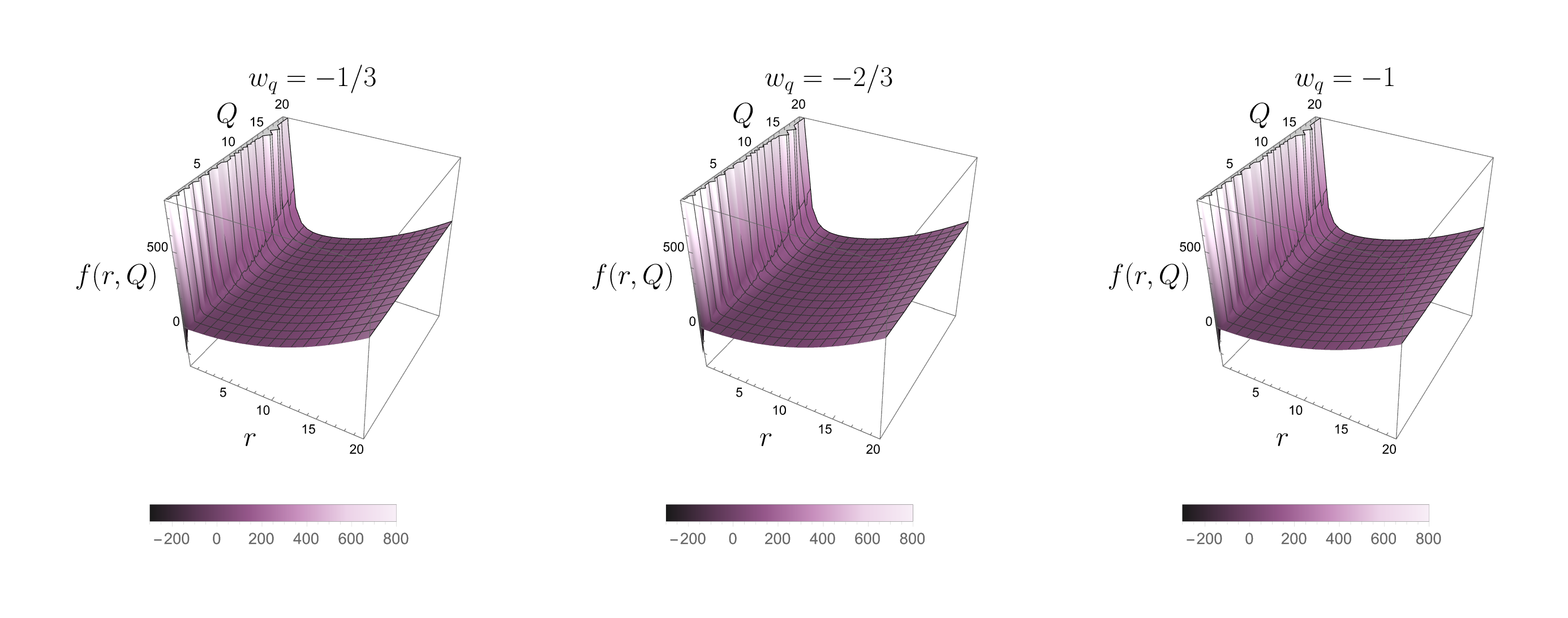}
        \caption{Plot of the event horizon with respect to the radial variable $r$ and the parameter $Q$ with $m=1$ and $N_q=-1$.}
        \label{f2}
\end{figure}

\begin{figure}[!htb]
     \centering
        \includegraphics[width=0.9\textwidth]{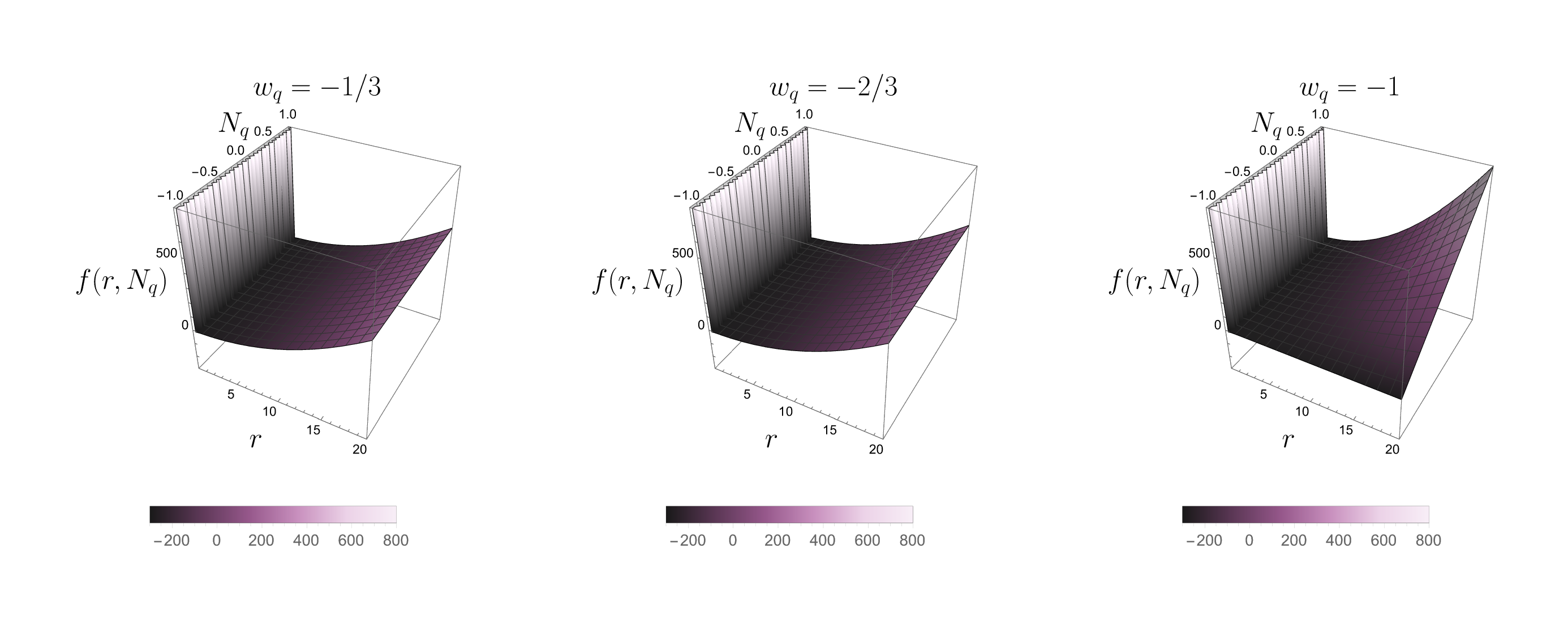}
        \caption{Plot of the event horizon with respect to the radial variable $r$ and the parameter $N_q$ with $m=1$ and $Q=1$.}
        \label{f3}
\end{figure}

\begin{figure}[!htb]
     \centering
        \includegraphics[width=0.9\textwidth]{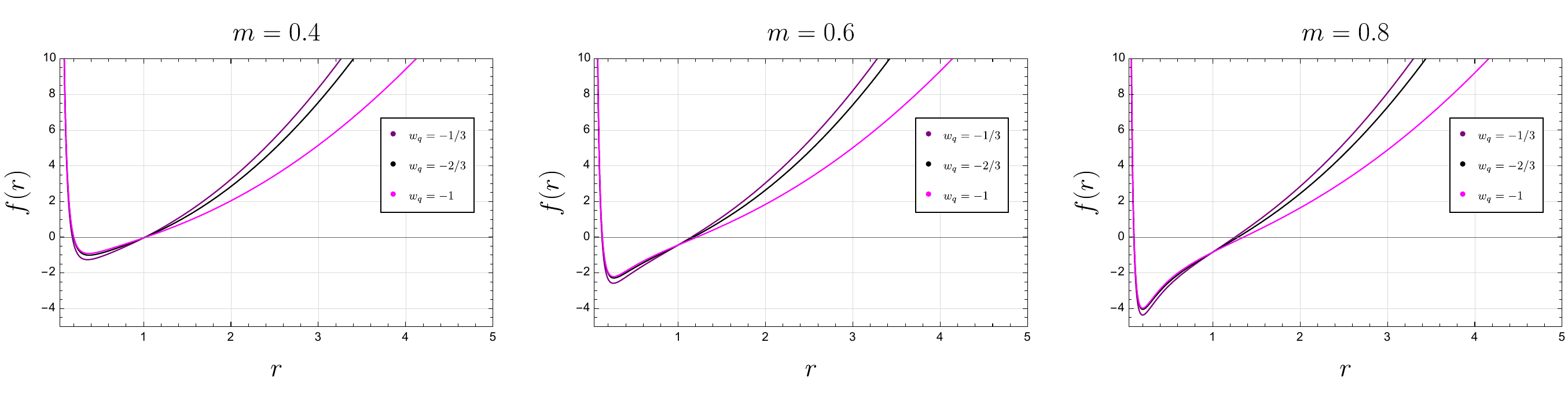}
        \caption{Plot of the event horizon with respect to the radial variable $r$. With $Q=0.4$ and $N_q=-0.4$.}
        \label{f4}
\end{figure}

\begin{figure}[!htb]
     \centering
        \includegraphics[width=0.9\textwidth]{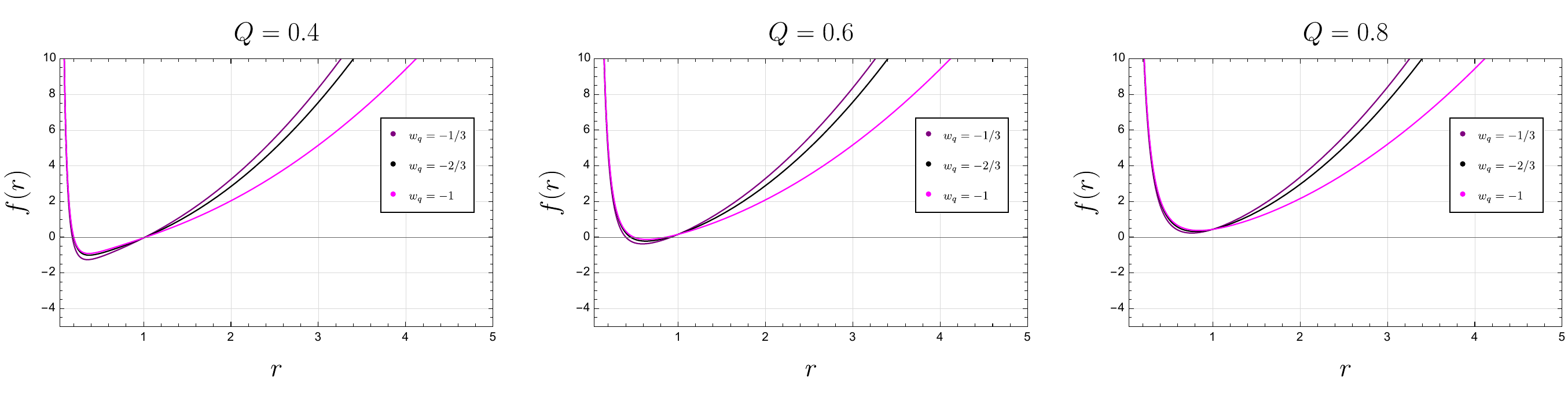}
        \caption{Plot of the event horizon with respect to the radial variable $r$. With $m=0.4$ and $N_q=-0.4$.}
        \label{f5}
\end{figure}

\begin{figure}[!htb]
     \centering
        \includegraphics[width=0.9\textwidth]{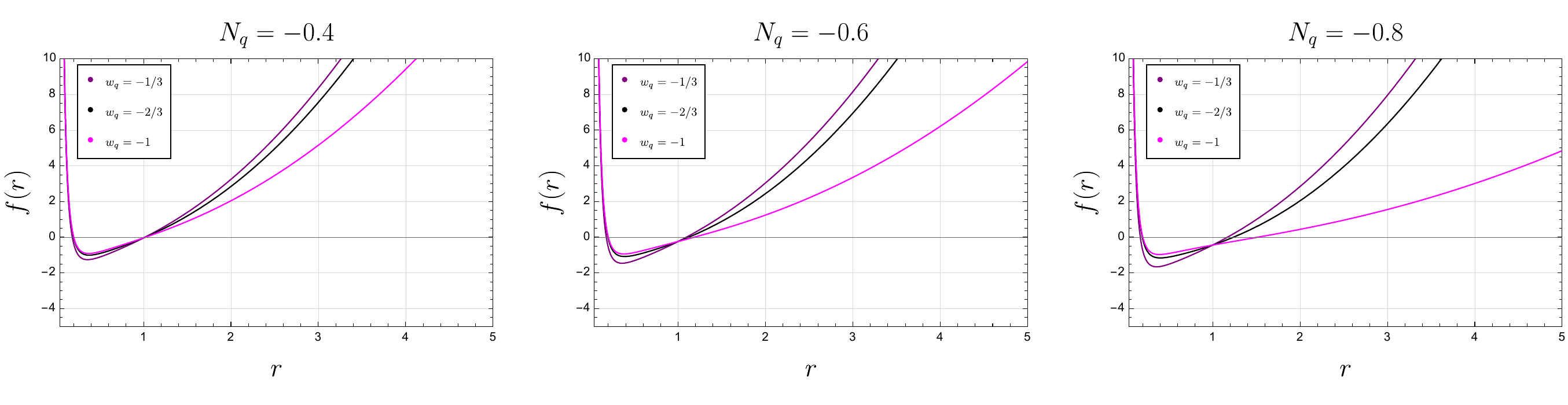}
        \caption{Plot of the event horizon with respect to the radial variable $r$. With $m=0.4$ and $Q=0.4$.}
        \label{f6}
\end{figure}

At this point, we can investigate the use of the particular form of the anisotropic fluid and the electromagnetic field on the right-hand side of the Einstein field equations.  
Searching for a solution with cylindrical symmetry, we take the following metric of the black string with the function $f\left(r\right)$ to be determined
\begin{equation}\label{MetricaCordaNegra}
ds^{2}=-f\left(r\right)dt^{2}+\frac{dr^{2}}{f\left(r\right)}+r^{2}d\phi^{2}+\frac{r^{2}}{l^{2}}dz^{2},
\end{equation}
where $-\infty<t<\infty$, $0\leq r<\infty$, $0\leq\phi<2\pi$ and $-\infty<z<\infty$, and by substituting Equation (\ref{MetricaCordaNegra}) in the Einstein field equations we obtain the following differential equations
\begin{equation}
\frac{1}{r}f^{'}\left(r\right)+\frac{1}{r^{2}}f\left(r\right)-\frac{3}{l^{2}}=-\left(h^{'}\left(r\right)\right)^{2}-8\pi\rho_{q},
\label{eq7}
\end{equation}
 \begin{equation}
\frac{1}{2}f^{''}\left(r\right)+\frac{1}{r}f^{'}\left(r\right)-\frac{3}{l^{2}}=\left(h^{'}\left(r\right)\right)^{2}+8\pi\rho_{q}\frac{1}{2}\left(3 w_{q}+1\right).
\label{eq8}
 \end{equation}
We should note that to solve these differential equations, it is necessary to find an explicit form for $h(r)$. In addition, $\rho = \rho(r)$, which means that the energy density can be determined directly by the field equations and must satisfy the equations (\ref{eq7}) and (\ref{eq8}).  Through the utilization of Maxwell's equations, we derive that $h\left(r\right)=\frac{lQ}{r}$, which allows us to write
\begin{equation}\label{Eqf(r)}
r^{2}f^{''}\left(r\right)+3rf^{'}\left(r\right)\left(w_{q}+1\right)+f\left(r\right)\left(3w_{q}+1\right)+\left(3w_{q}-1\right)\frac{l^{2}Q^{2}}{r^{2}}-9l^{-2}r^{2}\left(w_{q}+1\right)=0,
\end{equation}
which is a second order differential equation that carries information concerning the Einstein field equations. Solving the differential equation (\ref{Eqf(r)}) \textcolor{black}{via direct integration (For more details, see (\cite{luissantos}))}, we can write
\begin{equation}\label{SolI}
f\left(r\right)=\frac{r^{2}}{l^{2}}-\frac{2ml}{r}+\frac{N_{q}}{r^{3w_{q}+1}}+\frac{l^{2}Q^{2}}{r^{2}},
\end{equation}
where $m$ is the mass density, $Q$ the charge density and $N_{q}$ an integration constant \textcolor{black}{and by substituting this solution into field equations (\ref{eq7}) or (\ref{eq8}), the energy density $\rho_q$ associated with the Kiselev fluid can be written as
\begin{equation}
   \rho_q = \frac{3N_q w_q}{8\pi r^{3(w_q + 1) }}.  
\end{equation}
Thus we can see that the integration constant $N_q$ can be interpreted as a measure of the intensity of the energy density of the Kiselev fluid. As we will see in what follows, the value of the sign of $N_q$ is related to the violation of the weak and strong energy conditions. Equations (\ref{eq7}) and (\ref{eq8}) and the energy density $\rho_q$ suggest that the total energy density $\rho$ and the total pressures $p_\phi$ and $p_z$ are given by
\begin{equation}
    \rho =- p_{r} = \frac{3N_q w_q}{8\pi r^{3(w_q + 1) }} + \frac{l^2Q^2}{8\pi r^4} - \frac{3}{8\pi l^2},
    \label{p1}
\end{equation}
\begin{equation}
    p_\phi = p_z = \frac{3N_qw_q( 3w_q+1)}{16\pi r^{3(w_q + 1) }} + \frac{l^2Q^2}{8\pi r^4} + \frac{3}{8\pi l}.
    \label{p2}
\end{equation}
By using these expressions, we can obtain the energy conditions associated with solution (\ref{SolI}). The weak energy condition (WEC) and strong energy condition (SEC) are represented respectively by the equations
\begin{equation}
    \rho \geq 0,\:\:\rho + p_i \geq 0,\:\:\text{and} \:\:\rho + p_i \geq 0,\:\:\rho + \sum_{i} p_i \geq 0.
\end{equation}
Thus, substituting equations (\ref{p1}) and (\ref{p2}) into above energy conditions, we obtain the following expressions for both WEC and SEC
\begin{align}
     \frac{3N_q w_q}{8\pi r^{3(w_q + 1) }} + \frac{l^2Q^2}{8\pi r^4} - \frac{3}{8\pi l^2} \geq 0, \label{e1}\\
       \frac{3N_qw_q( 3w_q+1) + 6N_qw_q}{16\pi r^{3(w_q + 1) }} + \frac{l^2Q^2}{4\pi r^4}   \geq 0,\\
        \frac{3N_qw_q( 3w_q+1) }{8\pi r^{3(w_q + 1) }} + \frac{l^2Q^2}{4\pi r^4} +  \frac{3}{4\pi l^2} \geq 0.
\end{align}
To establish the conditions under which such equations are satisfied, we can set values for the parameters used. Initially choosing $l=Q=-N_q =1$, we have that for the value $w=-1$ the energy conditions are satisfied for any value of $r$. In contrast, if $N_q =l=1$ the SEC and WEC are satisfied only for $0< r \leq 0.6389 l.$ For $l=Q=1$ and $w=0$, the energy conditions are satisfied regardless of the value of $N_q$ if $0< r \leq 0.7598l$.} As we can see, expression (\ref{SolI}) allows the existence of event horizons, so it is interesting to investigate the dependence of these results on the parameters of the theory. In Figures \ref{f1}, \ref{f2}, and \ref{f3}, we can observe the horizon of events in the plot of $f(r)$ with respect to the parameters $m$, $Q$ and $N_q$, respectively. On the scale analyzed in these graphs, we can see that the sizes of the event horizons are very similar. In contrast, in Figures \ref{f4}, \ref{f5}   and \ref{f6}, we consider the graphs of $f(r)$ using three different values for $m, Q$ and $N_q$, respectively. As we can see in Figure \ref{f5}, the event horizon disappears for $Q=0.8$ and $w = -1,-2/3,-1/3$. In the Figures $N_{q}$ is given in units of $l^{3w_{q}+1}$ and $r$ in units of $l$.  

\textcolor{black}{To calculate the Kretschmann scalar for the metric given in eq. (\ref{MetricaCordaNegra}), taking into account the components of the Riemann tensor computed in the Appendix \ref{Riemann_Tensor_Components}, we have
\begin{equation}
  K=R_{\mu\nu\alpha\beta}R^{\mu\nu\alpha\beta}=\left(f^{''}\left(r\right)\right)^{2}+\frac{4}{r^{4}}\left(r^{2}\left(f^{'}\left(r\right)\right)^{2}+f^{2}\left(r\right)\right)
\end{equation}}
\textcolor{black}{and consequently, for the solution in (\ref{SolI}), the expression becomes}
 \begin{multline}
   R_{\mu\nu\rho\sigma}R^{\mu\nu\rho\sigma}=\frac{24}{l^{4}}+\frac{48m^{2}l^{3}}{r^{6}}-\frac{96l^{4}mQ^{2}}{r^{7}}+\frac{56l^{4}Q^{4}}{r^{8}}\\+\frac{12w_{q}\left(1+3w_{q}\right)N_{q}}{l^{2}r^{3w_{q}+3}}+\frac{3\left(27w_{q}^{4}+54w_{q}^{3}+51w_{q}^{2}+20w_{q}+4\right)N_{q}^{2}}{r^{6w_{q}+6}}\\-\frac{24(w_{q}+1)(3w_{q}+2)ml^{2}N_{q}}{r^{3w_{q}+6}}+\frac{12(w_{q}+1)(9w_{q}+4)N_{q}l^{2}Q^{2}}{r^{3w_{q}+7}}.
\end{multline} 
This expression is associated with the curvature of spacetime and reveals the existence of singularities in the solution found.
In the absence of charge and fluid, which means $Q = N_q = 0$, the above expression reduces to the Kretschmann scalar of the black string, see \cite{Lemos:1994xp}, furthermore, when $N_q = 0 $ the result is consistent with \cite{Lemos:1995cm}. We may also notice that the Kretschmann scalar obtained in \cite{Ahmed:2011qh} appears to be incompatible with the previous results, for $N_q = 0$ it vanishes. \textcolor{black}{It is interesting to determine the Kretschmann scalar in some particular cases, \( w_q = -1 \), \( w_q = -2/3 \), and \( w_q = -1/3 \). }

\textcolor{black}{For the case \( w_q = -1 \), it is given by
\begin{equation}
R_{\mu\nu\rho\sigma}R^{\mu\nu\rho\sigma}=\frac{24}{l^{4}}+\frac{48m^{2}l^{3}}{r^{6}}-\frac{96l^{4}mQ^{2}}{r^{7}}+\frac{56l^{4}Q^{4}}{r^{8}}+24N_{q}^{2}+\frac{24}{l^{2}}N_{q},
\end{equation}}

\textcolor{black}{for \( w_q = -2/3 \) it becomes
\begin{equation}
R_{\mu\nu\rho\sigma}R^{\mu\nu\rho\sigma}=\frac{24}{l^{4}}+\frac{48m^{2}l^{3}}{r^{6}}-\frac{96l^{4}mQ^{2}}{r^{7}}+\frac{56l^{4}Q^{4}}{r^{8}}+\frac{8N_{q}}{l^{2}r}+\frac{8N_{q}^{2}}{r^{2}}-\frac{8N_{q}l^{2}Q^{2}}{r^{5}}
\end{equation}}

\textcolor{black}{and finally, for \( w_q = -1/3 \),
\begin{equation}
R_{\mu\nu\rho\sigma}R^{\mu\nu\rho\sigma}=\frac{24}{l^{4}}+\frac{48m^{2}l^{3}}{r^{6}}-\frac{96l^{4}mQ^{2}}{r^{7}}+\frac{56l^{4}Q^{4}}{r^{8}}+\frac{4N_{q}^{2}}{r^{4}}-\frac{16ml^{2}N_{q}}{r^{5}}+\frac{8N_{q}l^{2}Q^{2}}{r^{6}}.
\end{equation}}
\textcolor{black}{These expressions show how the Kretschmann scalar varies with different values of the parameter \( w_q \), influencing the structure of spacetime singularities. }

The rotating counterpart of (\ref{SolI}) can be obtained by using the transformations \cite{Lemos:1994xp}
\begin{equation}
t\rightarrow\lambda t-a\phi,\qquad\phi\rightarrow\lambda\phi-\frac{a}{l^{2}}t,
\end{equation}
where $\lambda=\sqrt{1+\frac{a^{2}}{l^{2}}}$ and $a$ is the rotation parameter, and then we can explicitly write
\begin{equation}\label{SolII}
	\begin{split}
ds^{2}&=-\left(\frac{r^{2}}{l^{2}}-\frac{2ml}{r}+\frac{N_{q}}{r^{3w_{q}+1}}+\frac{l^{2}Q^{2}}{r^{2}}\right)\left(\lambda dt-ad\phi\right)^{2}+\frac{dr^{2}}{\left(\frac{r^{2}}{l^{2}}-\frac{2ml}{r}+\frac{N_{q}}{r^{3w_{q}+1}}+\frac{l^{2}Q^{2}}{r^{2}}\right)}\\&\qquad+\frac{r^{2}}{l^{4}}\left(adt-\lambda l^{2}d\phi\right)^{2}+\frac{r^{2}}{l^{2}}dz^{2},
	\end{split}
\end{equation}
\textcolor{black}{that is a metric which describes a charged rotating black string solution in the presence of quintessence, generalizing the Lemos black string by incorporating charge, rotation, and a quintessence field. As we can see, depending on the parameters of the model it presents event horizons, that is a subject that will be studied in the next section.}

\section{Event Horizons and Closed Time-like Curves}\label{secIII}

The event horizons of the solution (\ref{SolII}) can be obtained by the usual relation $g^{rr}=0$ which leads to
\begin{equation}\label{HorizonteEventos}
r^{3w_{q}+3}-2ml^{3}r^{3w_{q}}+r^{3w_{q}-1}l^{4}Q^{2}+l^{2}N_{q}=0.
\end{equation} 
In principle, the number of horizons of this metric is associated with the equation of state parameter $w_q$. Furthermore, the particular shape of the horizon depends on the charge $Q$ and the radius of curvature $l$. Thus, we should study particular values for $w_q$. 
We consider three cases of interest, $w_q=$ -1, -2/3 and -1/3.

    
    For $w_q = -1$, equation (\ref{HorizonteEventos}) reduces to the event horizon expression in the form   
    \begin{equation}
      r^{4}-\frac{2ml^{3}}{1+l^{2}N_{q}}r+\frac{l^{4}Q^{2}}{1+l^{2}N_{q}}=0,
      \label{26}
    \end{equation} 
        which is a polynomial equation of the fourth degree where the solutions give us the horizons. Due to the singularity at $l^2 N_q= -1$, this point must be excluded from Equation (\ref{26}).    \textcolor{black}{Solving this quartic equation, the positions of the inner and outer event horizons are obtained}
    \textcolor{black}{\begin{equation}
       r_{\pm}=\frac{\sqrt{\kappa}}{2}\pm\frac{1}{2}\sqrt{-\kappa+\frac{1}{\sqrt{\kappa}}\frac{4ml^{3}}{1+l^{2}N_{q}}}, 
    \end{equation} 
    with the constant $\kappa$ being the solution of the associated cubic equation,
    \begin{equation}
      \kappa=\sqrt[3]{q+\sqrt{p}}+\sqrt[3]{q-\sqrt{p}}
    \end{equation}
    \begin{equation}
        q=\frac{2m^{2}l^{6}}{\left(1+l^{2}N_{q}\right)^{2}},
    \end{equation}
    \begin{equation}
        p=\frac{4m^{4}l^{12}}{\left(1+l^{2}N_{q}\right)^{4}}-\frac{64l^{12}Q^{6}}{27\left(1+l^{2}N_{q}\right)^{3}},
    \end{equation}}
each term of these equations having a singularity at $l^2 N_q = -1 $.  
    
    For $w_q = -2/3$ equation (\ref{HorizonteEventos}) reduces to 
    \begin{equation}
       r^{4}-2ml^{3}r+l^{2}N_{q}r^{3}+l^{4}Q^{2}=0 ,
    \end{equation} 
    then the solution of this quartic equation determines the \textcolor{black}{horizons} through the relation
    \textcolor{black}{\begin{equation}
       r_{\pm}=\frac{\sqrt{\kappa}}{2}-\frac{1}{4}l^{2}N_{q}\pm\frac{1}{2}\sqrt{\frac{16ml^{3}-l^{6}N_{q}^{3}}{4\sqrt{\kappa}}+\frac{3}{4}l^{4}N_{q}^{2}-\kappa},
    \end{equation}}
   \textcolor{black}{ whose associated cubic equation gives 
     \begin{equation}
       \kappa=\frac{l^{4}N_{q}^{2}}{4}+\sqrt[3]{q+\sqrt{p}}+\sqrt[3]{q-\sqrt{p}},
    \end{equation}
    \begin{equation}
        q=\frac{l^{6}}{2}\left(l^{2}N_{q}^{2}Q^{2}+4m^{2}\right),
    \end{equation}
    \begin{equation}
        p=\frac{l^{12}}{4}\left(l^{2}N_{q}^{2}Q^{2}+4m^{2}\right)^{2}-\frac{8}{27}l^{12}\left(lmN_{q}+2Q^{2}\right)^{3},
    \end{equation}}
contrasting the last case, in this expression there is no singularity in $\kappa$ at $l^2 N_q  =-1$. 

Finally, for $w_q = -1/3$ the relevant equation for the horizon is 
        \begin{equation}
        r^{4}+l^{2}N_{q}r^{2}-2ml^{3}r+l^{4}Q^{2}=0,
        \label{31}
    \end{equation} 
    \textcolor{black}{whose solution for the event horizon is given as}
    \textcolor{black}{\begin{equation}
       r_{\pm}=\frac{\sqrt{\kappa}}{2}\pm\frac{1}{2}\sqrt{\frac{4ml^{3}}{\sqrt{\kappa}}-2l^{2}N_{q}-\kappa},
    \end{equation}}
    where we have considered the constants 
   \textcolor{black}{ \begin{equation}
       \kappa=\sqrt[3]{-q+\sqrt{p}}+\sqrt[3]{-q-\sqrt{p}}-\frac{2}{3}l^{2}N_{q},
    \end{equation} \begin{equation}
       q=-\frac{1}{27}l^{6}\left(54m^{2}+N_{q}^{3}-36N_{q}Q^{2}\right),
    \end{equation} \begin{equation}
       p=\frac{l^{12}}{729}\left(\left(54m^{2}+N_{q}^{3}-36N_{q}Q^{2}\right)^{2}-\left(N_{q}^{2}+12Q^{2}\right)^{3}\right).
    \end{equation}}
These are the horizons associated with some particular values of $w_q$. The values obtained for $r_{\pm}$ in this section are important for studying the thermodynamics of black holes since, at the horizon of this type of astrophysical object, we have several classical and quantum physical phenomena occurring. \textcolor{black}{As we can see from the equations which define the horizons (\ref{HorizonteEventos}), the structure of spacetime is composed by the string in addition to two elements which permeate space: the quintessence fluid and the electric field. Thus, the charged black string described by this formulation exhibits a complex horizon structure influenced by both the electric charge and the nature of the surrounding fluid.} 

\textcolor{black}{One interesting scenario occurs when \( r_+ = r_- \), which characterizes extremal solutions, important for studying the nature of singularities and their implications. By numerically analyzing the extremal solutions for \( w_q = -1, -2/3, -1/3 \) with \( m = 1 = l \), we obtain the results presented in Table \ref{tab:values}. The table indicates that the extremal charges decrease as \( N_q \) approaches zero, irrespective of the value of \( w_q \). Extreme values (\( N_q = -1, w_q = -1 \)) suggest a divergence, while for \( N_q = 0 \), the charges converge to a fixed value. More negative values of \( w_q \) lead to slightly higher charges. 
In particular, for the case \( w_q = -1 \) and $N_q > -1$, we find that the extremal charges are given by the expression:
\begin{align}
Q_{\text{ext}} = 3^{\frac{1}{2}}\left(\frac{1}{16N_q + 16}\right)^{\frac{1}{6}}.
\end{align}
This relation highlights the sensitivity of the extremal charge to the parameter \( N_q \). As \( N_q \to -1 \), the charge diverges, in line with the numerical observations. Conversely, as \( N_q \to 0 \), the charge approaches a finite value, matching the convergence seen in Table \ref{tab:values}. Furthermore, it suggests that the interplay between the quintessence field and the charge is highly non-linear, despite the fact that small changes in \( N_q \) in the numerical analysis are not leading to significant variations in the extremal charge.
}

\begin{table}[H]
\centering
\begin{tabular}{c|ccccccccc}
\hline
\(w_{q} \backslash N_{q}\) & \(-1.00\) & \(-0.75\) & \(-0.5\) & \(-0.25\) & \(0\) & $0.25$ & $0.5$ & $0.75$ & $1.00$\\ \hline
\(-1\) & \(\infty\) & \(1.38\) & \(1.22\) & \(1.14\) & \(1.09\) & $1.05$ & $1.02$ & $0.99$ & $0.97$\\
\(-\frac{2}{3}\) & \(1.44\) & \(1.32\) & \(1.23\) & \(1.15\) & \(1.09\) & \(1.04\) & \(0.99\) & \(0.96\) & \(0.92\) \\
\(\frac{1}{3}\) & \(1.41\) & \(1.33\) & \(1.24\) & \(1.17\) & \(1.09\) & \(1.02\) & \(0.96\) & \(0.90\) & \(0.84\) \\ \hline
\end{tabular}
\caption{Extremal charges $Q_{\text{ext}}$} as a function of $N_q$ for different values of $w_q$ and parameters $m = l =1$.
\label{tab:values}
\end{table}

\textcolor{black}{An interesting aspect of this kind of spacetime are the closed timelike curves (CTCs), which arise in rotating charged black string spacetimes when the angular metric component satisfies \( g_{\phi\phi} < 0 \), signaling local causality violation. So, for the line element considered above, the condition \( g_{\phi\phi} = 0 \) yields the critical radius \( r_{\text{CTC}} \), governed by
\begin{equation}\label{CTC}
l^{2}r^{3w_{q}+3}+2ml^{3}a^{2}r^{3w_{q}}-l^{4}Q^{2}a^{2}r^{3w_{q}-1}-N_{q}l^{2}a^{2}=0,
\end{equation}
and is determined by the parameters of the black string, specifically the mass \( m \), the AdS curvature radius \( l \), the quintessence density \( N_q \), the charge \( Q \), and the equation-of-state parameter \( w_q \). These parameters, along with the rotation parameter \( a \), define the structure of the solution. Below, we analyze \( r_{\text{CTC}} \) for some specific values of \( w_q \).}

\textcolor{black}{For \(w_q = -1\), Eq.~(\ref{CTC}) reduces to the quartic equation
\begin{equation}
  r^{4}-\frac{2mla^{2}}{\left(N_{q}a^{2}-1\right)}r+\frac{l^{2}Q^{2}a^{2}}{\left(N_{q}a^{2}-1\right)}=0,
\end{equation}
and its physical solution determines the CTC radius
\begin{equation}
  r_{\text{CTC}}=\frac{1}{2}\sqrt{-\kappa+\frac{4mla^{2}}{\sqrt{\kappa}\left(N_{q}a^{2}-1\right)}}-\frac{1}{2}\sqrt{\kappa},
\end{equation}
where \(\kappa\) is defined via the resolvent cubic
\begin{equation}
    \kappa=\sqrt[3]{\Theta+\sqrt{\Xi}}+\sqrt[3]{\Theta-\sqrt{\Xi}},
\end{equation}
\begin{equation}
    \Theta=\frac{2m^{2}l^{2}a^{4}}{\left(a^{2}N_{q}-1\right)^{2}},
\end{equation}
\begin{equation}
    \Xi=\frac{4a^{8}l^{4}m^{4}}{\left(a^{2}N_{q}-1\right)^{4}}-\frac{64a^{6}l^{6}Q^{6}}{27\left(a^{2}N_{q}-1\right)^{3}}.
\end{equation}}

\textcolor{black}{For \(w_q = -2/3\), the constraint becomes
\begin{equation}
   r^{4}-N_{q}a^{2}r^{3}+2mla^{2}r-l^{2}Q^{2}a^{2}=0,
\end{equation}
and the CTC radius is then obtained by solving the quartic,
\begin{equation}
  r_{\text{CTC}}=\frac{1}{2}\sqrt{-\kappa-\frac{3}{4}N_{q}^{2}a^{4}+\frac{\frac{1}{4}N_{q}^{3}a^{6}-4mla^{2}}{\sqrt{\kappa}}}-\frac{1}{2}\sqrt{\kappa}+\frac{N_{q}a^{2}}{4},
\end{equation}
with the parameter $\kappa$ being the solution of the associated cubic, and given by
\begin{equation}
    \kappa=\sqrt[3]{\Theta+\sqrt{\Xi}}+\sqrt[3]{\Theta-\sqrt{\Xi}}-\frac{1}{4}N_{q}^{2}a^{4},
\end{equation}
\begin{equation}
    \Theta=-\frac{1}{2}a^{4}l^{2}\left(4m^{2}-a^{2}N_{q}^{2}Q^{2}\right),
\end{equation}
\begin{equation}
    \Xi=\frac{1}{4}a^{8}l^{4}\left(a^{2}N_{q}^{2}Q^{2}-4m^{2}\right)^{2}-\frac{8}{27}a^{6}l^{3}\left(a^{2}mN_{q}-2lQ^{2}\right)^{3}.
\end{equation}}

\textcolor{black}{Finally, for \(w_q = -1/3\), Eq.~(\ref{CTC}) simplifies to
\begin{equation}
   r^{4}-N_{q}a^{2}r^{2}+2mla^{2}r-l^{2}Q^{2}a^{2}=0
\end{equation}
and the corresponding solution is
\begin{equation}
  r_{\text{CTC}}=\frac{1}{2}\sqrt{-\kappa-2N_{q}a^{2}-\frac{4mla^{2}}{\sqrt{\kappa}}}-\frac{1}{2}\sqrt{\kappa},
\end{equation}
where the unknown parameters are
\begin{equation}
    \kappa=\frac{2}{3}N_{q}a^{2}+\sqrt[3]{\Theta+\sqrt{\Xi}}+\sqrt[3]{\Theta-\sqrt{\Xi}},
\end{equation}
\begin{equation}
    \Theta=-\frac{1}{27}a^{4}\left(a^{2}N_{q}^{3}+l^{2}\left(36N_{q}Q^{2}-54m^{2}\right)\right),
\end{equation}
\begin{equation}
    \Xi=\frac{1}{729}\left(a^{8}\left(a^{2}N_{q}^{3}+l^{2}\left(36N_{q}Q^{2}-54m^{2}\right)\right)^{2}-\left(a^{4}N_{q}^{2}-12a^{2}l^{2}Q^{2}\right)^{3}\right). 
\end{equation}
Analyzing these three scenarios, we observe the existence of the solutions, i.e. the critical radii presented above, occurs up to the positivity of the respective discriminants of the quartic equations}. \textcolor{black}{In principle, this positivity requirement translates into a condition on the rotating parameter $a$. In fact, for the $w_q = - 1$ case, we encountered two scenarios. When $N_q <432m^{4}/(256l^{2}Q^{6})$, the condition on $a$ reads \begin{align}
    a^{2}<\frac{256l^{2}Q^{6}}{256l^{2}Q^{6}N_{q}-432m^{4}}.
\end{align} Nevertheless, this inequality is never satisfied for real values of $a$, implying that the discriminant remains positive, suggesting that closed timelike curves (CTCs) cannot arise, as the solutions for the critical radius, $r_{CTC}$, are purely imaginary. Conversely, in the regime where $N_q > 432m^{4}/(256l^{2}Q^{6})$, the inequality sign is reversed, i.e. \begin{align}
    a^{2}>-\frac{256l^{2}Q^{6}}{\left| 256l^{2}Q^{6}N_{q}-432m^{4} \right|},
\end{align} potentially yielding a positive discriminant. In this case, a positive real solution for the critical radius becomes possible. For the remaining scenarios, specifically $w_q = -2/3$ and $w_q = -1/3$, the complexity of the discriminant precludes an analytical determination of conditions on the parameter $a$. A numerical investigation with representative values $m = l = 1$ and $Q = 0.1$ was conducted, revealing no real solutions for the critical radii equations within the range $-1 < N_q < 1$. These findings indicate that the CTCs may be present but they are contingent on specific parameter choices on the parameter space.}

 \textcolor{black}{An intriguing feature of the rotating solution 
 (\ref{SolII}) is the appearance of an angular deficit, as the ratio between a proper circumference and its radius differs from $2\pi$. In fact, considering the metric (\ref{SolHorizon_2D}) at the external horizon on the angular sector, i.e. $ds^2_\chi = \lambda^2 r_+^2 d\chi^2$, the proper length of the circumference of radius $r_+$ is \begin{align}
     \mathcal{C} = \int_{0}^{2\pi}\sqrt{g_{\chi\chi}}d\chi=2\pi\lambda r_{+}.
 \end{align} Therefore, the angular deficit reads \begin{align}
     \delta=\frac{\mathcal{C}_0-\mathcal{C}}{r_{+}}=2\pi\left(1-\lambda\right),
 \end{align} where $\mathcal{C}_0$ is the the proper length of the circumference for $\lambda = 1$, i.e. $a = 0$. The angular deficit obtained does not have any dependence on the parameters of the solution (\ref{SolI}), and it vanishes for the nonrotating case.
 }

\section{Mass, angular momentum and charge}\label{secIV}
\textcolor{black}{Another aspect of the presented solution lies in its conserved charges. We employ the Brown-York formalism \cite{brown1993quasilocal} for quasilocal conserved charges to analyze the solution (\ref{SolII}), calculating the quasilocal mass, angular momentum, and charge associated with the black string in the presence of a quintessence fluid. We start by considering the metric (\ref{SolII}) in the canonical form 
\begin{align}\label{Canonical}
    ds^{2}&=-\left(N^{0}\right)^{2}dt^{2}+R^{2}\left(N^{\phi}dt+d\phi\right)^{2}+\frac{dr^{2}}{f(r)}+e^{-4\varphi}dz^{2},
\end{align}where the coefficient functions are given by 
\begin{align}
    \left(N^{0}\right)^{2}& =\left(\lambda^{2}-\frac{a^{2}}{l^{2}}\right)^{2}\left(\frac{r^{2}}{l^{2}}-\frac{2ml}{r}+\frac{l^{2}Q^{2}}{r^{2}}+\frac{N_{q}}{r^{3w_{q}+1}}\right)\frac{r^{2}}{R^{2}}, \\
    R^{2}& =\lambda^{2}r^{2}-a^{2}\left(\frac{r^{2}}{l^{2}}-\frac{2ml}{r}+\frac{l^{2}Q^{2}}{r^{2}}+\frac{N_{q}}{r^{3w_{q}+1}}\right),\\
    N^{\phi}&=-\frac{\lambda a}{R^{2}}\left(\frac{2ml}{r}-\frac{l^{2}Q^{2}}{r^{2}}-\frac{N_{q}}{r^{3w_{q}+1}}\right), \quad
    e^{-4\varphi}=\frac{r^{2}}{l^{2}}.
\end{align} 
The functions $N^0$ and $N^\phi$ are the lapse and shift functions, respectively. We notice that for $N_q = 0$, these functions coincide with the results in \cite{Lemos:1995cm}. To compute the conserved charges, we consider a region of spacetime bounded by $r=const.$ and two spacelike hypersurfaces $t=t_1$ and $t=t_2$. Furthermore, as argued in \cite{Lemos:1995cm}, one must additionally restrict the calculations to regions between $z=z_1$ and $z=z_2$, since the black string extends along the entire $z$-axis, causing the expressions to diverge. This restricted region will be referred to as $B_z$, and the spacelike hypersurface of constant $t$ as $\Sigma$. Let $\sigma_{ab}$ be the metric obtained from (\ref{Canonical}) by setting $dt = dr =0$, while the three-space metric $h_{ij}$ is obtained by setting $dt=0$. Hence, the presence of two Killing vector fields $\xi_t = \partial / \partial t $ and $\xi_\phi = \partial / \partial \phi  $, allow us to define the charges \begin{align}\label{eq:mass_and_momentum}
\mathcal{Q}_{\xi}=\dfrac{1}{\Delta z}\int_{B_{z}}d^{2}x\sqrt{\sigma}\left(\epsilon u^{\mu}+j^{\mu}\right)\xi_{\mu},
\end{align} where $u^\mu$ is the timelike future directed normal to $\Sigma$, $\sigma$ is the determinant, $\epsilon$ is the energy surface density on $B_z$, and $j^\mu = (0,j^i)$ is the momentum surface density on $B_z$ given by \begin{align}
    \epsilon=\frac{k}{8\pi}, \qquad
    j^{i}=\frac{\sigma_{j}^{i}n_{k}\Pi^{jk}}{16\pi\sqrt{h}},
\end{align} being $n^k$ the unit vector normal to $B_z$ on $\Sigma$, $k$ the trace of the extrinsic curvature of $B_z$, $\Pi^{jk}$ the conjugate momentum in $\Sigma$, and $h$ the determinant of $h_{ij}$. As for the charge of the solution, it can be obtained by \begin{align}\label{eq:charge}
    \mathcal{Q} = \dfrac{1}{4\pi\Delta z} \int_{B_z} d^2x \dfrac{n_i\mathcal{E}^i\sqrt{\sigma}}{\sqrt{h}}
\end{align} where $\mathcal{E}^i$ is the canonical conjugate momentum of $A_\mu dx^\mu$. 
\textcolor{black}{To obtain finite results at the asymptotic boundary, it is necessary to subtract the AdS background contribution. The expressions must be regularized to remove divergences arising at spatial infinity, which is done by subtracting the background values of the extrinsic curvature and the conjugate momenta \cite{Lemos:1995cm}. The regularized quantities are defined as
\begin{equation}
k_{\text{reg}} = k - k_0, \quad j^i_{\text{reg}} = j^i - j^i_0, \quad \mathcal{E}^i_{\text{reg}} = \mathcal{E}^i - \mathcal{E}^i_0,
\end{equation}
where \(k\) is the trace of the extrinsic curvature, \(j^i\) the gravitational conjugate momentum, and \(\mathcal{E}^i\) the electromagnetic conjugate momentum. The subscript \(0\) refers to background values obtained by setting \(m = Q = N_q = 0\).} After this procedure, we obtain the mass, angular momentum and charge as 
\begin{equation}\label{charges}
    \mathcal{M}=\frac{1}{4}\left(3\lambda^{2}-1\right)m,\quad\mathcal{J}=\frac{3}{4}\lambda ma,\quad\mathcal{Q}=\frac{\lambda Q}{2}.
\end{equation}
The physical parameters of the charged and rotating black string are then influenced by the rotation parameter \( a \) through the dependence of \(\lambda(a)\). 
\textcolor{black}{In particular, as shown in Sec.~\ref{secVI}, the mass parameter \( m \) can be expressed in terms of the solution parameters at the event horizon through Eq.~(\ref{eq:parameter_m}). This relation makes explicit the dependence of the conserved charges on \( a \), \( N_q \), \( Q \), \( l \), \( r_+ \), and \( w_q \).
}
The mass, angular momentum, and charge are modified by this relationship, showing how rotation affects the system. Furthermore, the results are compatible with ones obtained in \cite{Lemos:1995cm,Ali:2019mxs, Bakhtiarizadeh:2023mhk}.
} 

\textcolor{black}{Based on the conserved charges given in Eq.~\eqref{charges}, the first law of thermodynamics for this system can be expressed as
\begin{equation}
    d\mathcal{M} = T_{+} \, dS_{+} + \Omega_{+} \, d\mathcal{J} + \Theta_{q} \, dN_{q} + \Phi_{+} \, d\mathcal{Q},
\end{equation}
where the conjugate thermodynamic variables are defined by the following partial derivatives:
\begin{equation}
    T_{+} = \left( \frac{\partial \mathcal{M}}{\partial S_{+}} \right)_{\mathcal{J}, N_q, \mathcal{Q}}, \quad
    \Omega_{+} = \left( \frac{\partial \mathcal{M}}{\partial \mathcal{J}} \right)_{S_{+}, N_q, \mathcal{Q}},
\end{equation}
\begin{equation}
    \Theta_{q} = \left( \frac{\partial \mathcal{M}}{\partial N_{q}} \right)_{S_{+}, \mathcal{J}, \mathcal{Q}}, \quad
    \Phi_{+} = \left( \frac{\partial \mathcal{M}}{\partial \mathcal{Q}} \right)_{S_{+}, \mathcal{J}, N_q}.
\end{equation}
Here, \( T_+ \), \( \Omega_+ \), \( \Theta_q \), and \( \Phi_+ \) represent, respectively, the temperature, angular velocity, quintessence potential, and electric potential evaluated at the event horizon \cite{Lemos:1995cm,Ali:2019mxs, Bakhtiarizadeh:2023mhk,Dehghani2002}.
}

We are also interested in determining the temperature associated with particle formation due to quantum processes in the horizon region. In the next section, we will analyze this effect in detail.

\section{Scalar particles from charged rotating black strings}\label{secV}

To study the contribution of charged scalar particles to Hawking radiation from charged rotating black strings immersed in a Kiselev anisotropic fluid, we will represent these particles by the Klein-Gordon equation written in the metric derived in Sec. II,
\begin{equation}\label{EqKGTunelamento}
	\frac{1}{\sqrt{-g}}\left(\partial_{\alpha}-\frac{ie}{\hslash}A_{\alpha}\right)\left[\sqrt{-g}g^{\alpha\beta}\left(\partial_{\beta}-\frac{ie}{\hslash}A_{\beta}\right)\Psi\right]-\frac{m^{2}}{\hslash^{2}}\Psi=0,
\end{equation}
where $e$ and $m$ are the charge and mass of the scalar particles. We apply the WKB approximation and assume an ansatz of the form
\textcolor{black}{\begin{equation}
   \Psi\left(t,r,\phi,z\right)=\exp\left[\frac{i}{\hslash}I\left(t,r,\phi,z\right)\right],
\end{equation}}
and then, to the leading order in $\hslash$, we obtain the relativistic Hamilton–Jacobi equation
\begin{equation}    g^{\alpha\beta}\left(\partial_{\alpha}I\partial_{\beta}I+e^{2}A_{\alpha}A_{\beta}-2eA_{\alpha}\partial_{\beta}I\right)+m^{2}=0 .
\end{equation}
Replacing the coefficients of (\ref{DiagMetric}) and carrying out some manipulations, we obtain
\textcolor{black}{\begin{equation} \label{eq001}
g^{tt}\left(\partial_{t}I-eA_{t}\right)^{2}+g^{rr}\left[\partial_{r}I\right]^{2}+g^{\chi\chi}\left(\partial_{\chi}I-eA_{\chi}\right)^{2}+g^{zz}\left[\partial_{z}I\right]^{2}+m^{2}=0.
\end{equation}}

Considering the spacetime symmetries of the background metric we assume the
following form of the solution of the above equation
\textcolor{black}{\begin{equation}\label{I(r)}
	I=-\left(E-J_{1}\Omega_+\right)t+W\left(r\right)+J_{1}\chi+J_{2}z,
\end{equation}}
where $E$, $J_{1}$ and $J_{2}$ are respectively the quantum numbers associated with energy, angular momentum and momentum in the $z$ direction, the angular velocity is given by 
\begin{equation}\label{angularvel}
    \Omega_{+} = \frac{a}{\lambda l^{2}}.
\end{equation} 

Replacing (\ref{I(r)}) in (\ref{eq001}) and solving for $W\left(r\right)$, we obtain the expression
\textcolor{black}{\begin{equation}
	W\left(r\right)=\pm\int dr\frac{\sqrt{\left(\left(\left[E-J_{1}\Omega\right]+eA_{t}\right)^{2}-\frac{F\left(r\right)}{H\left(r\right)}\left[J_{1}-eA_{\chi}\right]^{2}-F\left(r\right)\left[\frac{J_{2}^{2}}{K\left(r\right)}+m^{2}\right]\right)}}{\sqrt{F\left(r\right)G\left(r\right)}},
\end{equation}}
and integrating the above equation using the residue theory we get
\textcolor{black}{\begin{equation}
W\left(r\right)=\pm\frac{i\pi\left(\left[E-J_{1}\Omega_{+}\right]+eA_{t}\left(r_{+}\right)\right)}{\sqrt{F^{'}\left(r_{+}\right)G^{'}\left(r_{+}\right)}}=\pm\frac{i\pi\lambda\left(E-J_{1}\Omega_{+}+eA_{t}\left(r_{+}\right)\right)}{f^{'}\left(r_{+}\right)}.
\label{W+}
\end{equation}}

\begin{figure}[!htb]
     \centering
       \includegraphics[width=0.9\textwidth]{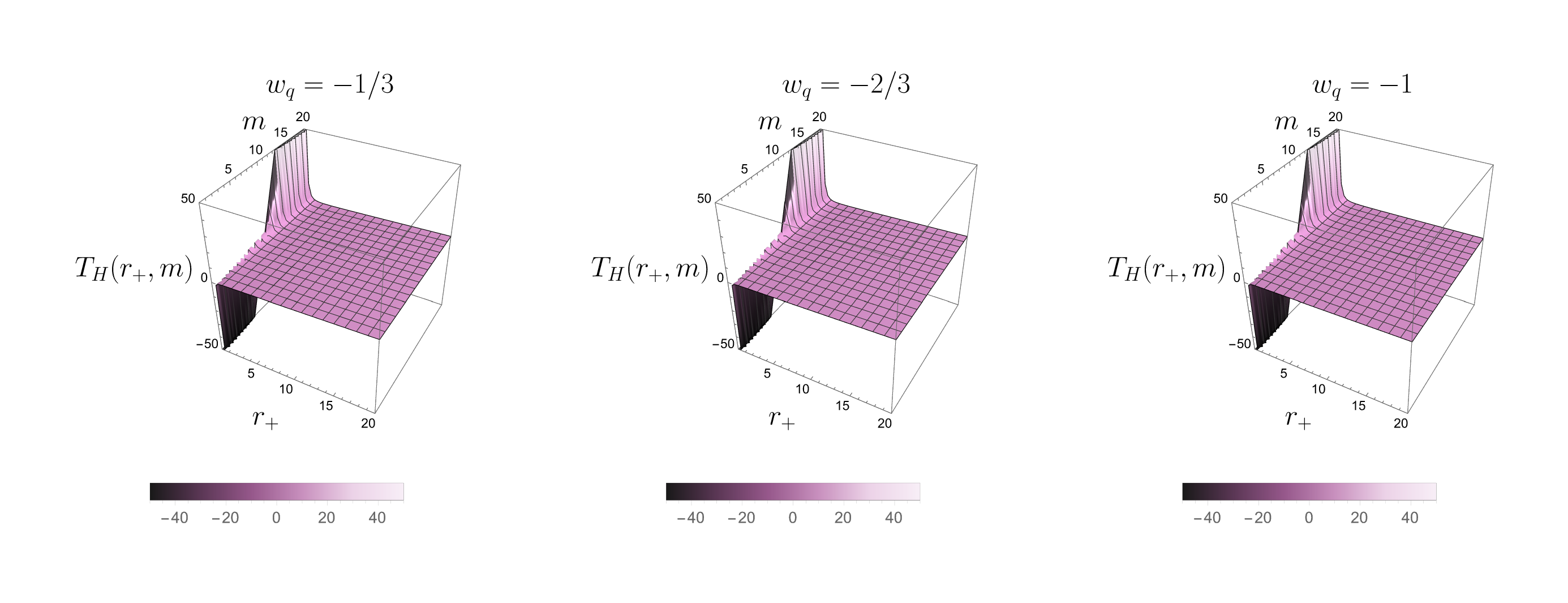}
        \caption{Plot of temperature versus event horizon radius $r_+$ and parameter $m$ with $Q=1$ and $N_q=-1$. $N_{q}$ is given in units of $l^{3w_{q}+1}$ and $r_+$ in units of $l$.}
        \label{f7}
\end{figure}

\begin{figure}[!htb]
     \centering
       \includegraphics[width=0.9\textwidth]{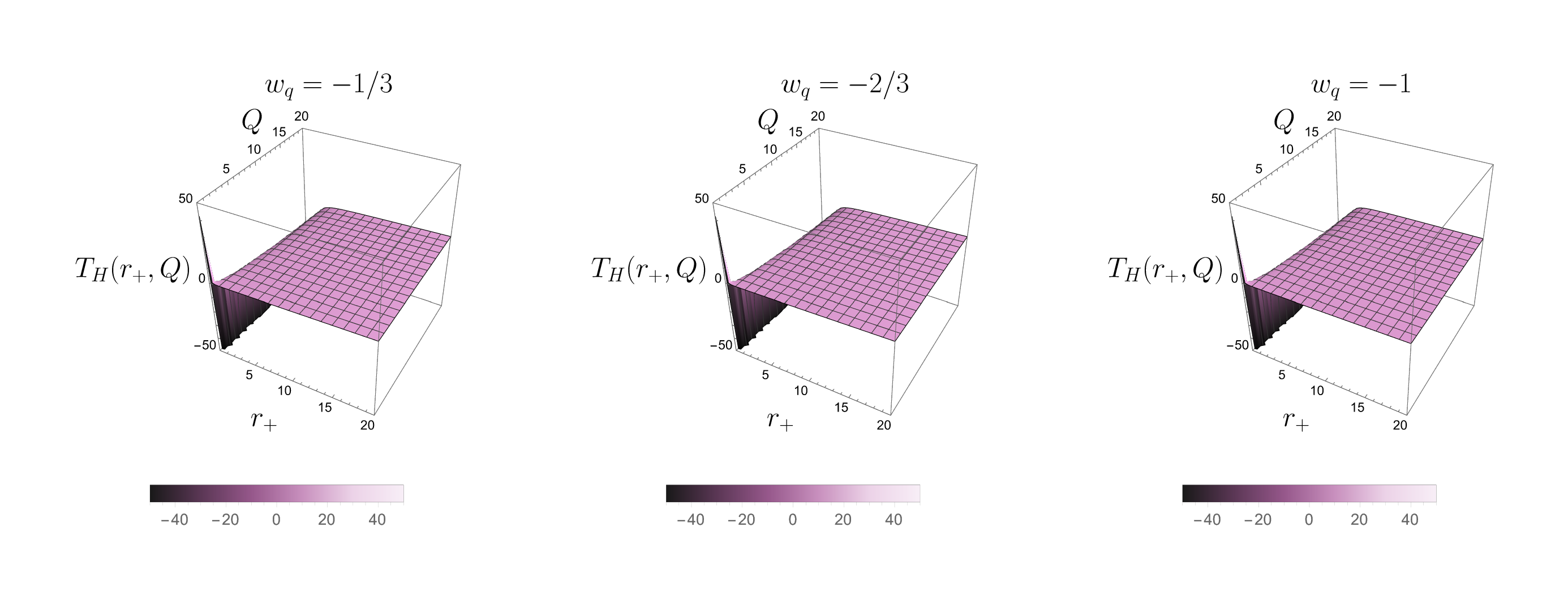}
        \caption{Plot of temperature versus event horizon radius $r_+$ and parameter $Q$ with $m=1$ and $N_q=-1$.}
        \label{f8}
\end{figure}

\begin{figure}[!htb]
     \centering
       \includegraphics[width=0.9\textwidth]{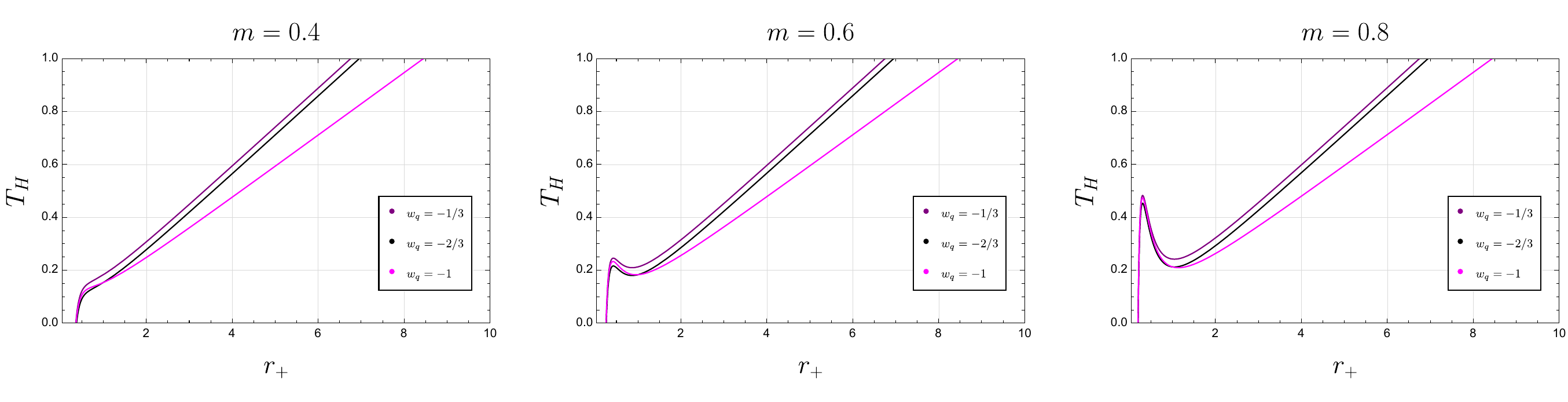}
      \caption{Plot of temperature versus event horizon radius $r_+$. With $Q=0.4$, $a/l=0.4$ and $N_q=-0.4$.}
        \label{f9}
\end{figure}

\begin{figure}[!htb]
     \centering
       \includegraphics[width=0.9\textwidth]{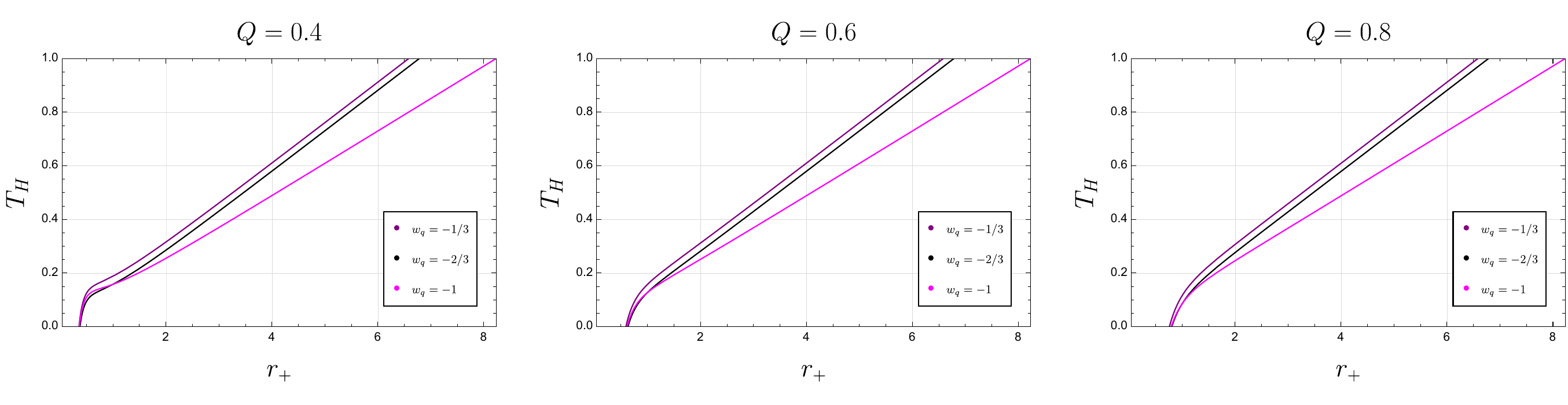}
      \caption{Plot of temperature versus event horizon radius $r_+$. With $m=0.4$, $a/l=0.4$ and $N_q=-0.4$.}
        \label{f10}
\end{figure}

\begin{figure}[!htb]
     \centering
       \includegraphics[width=0.9\textwidth]{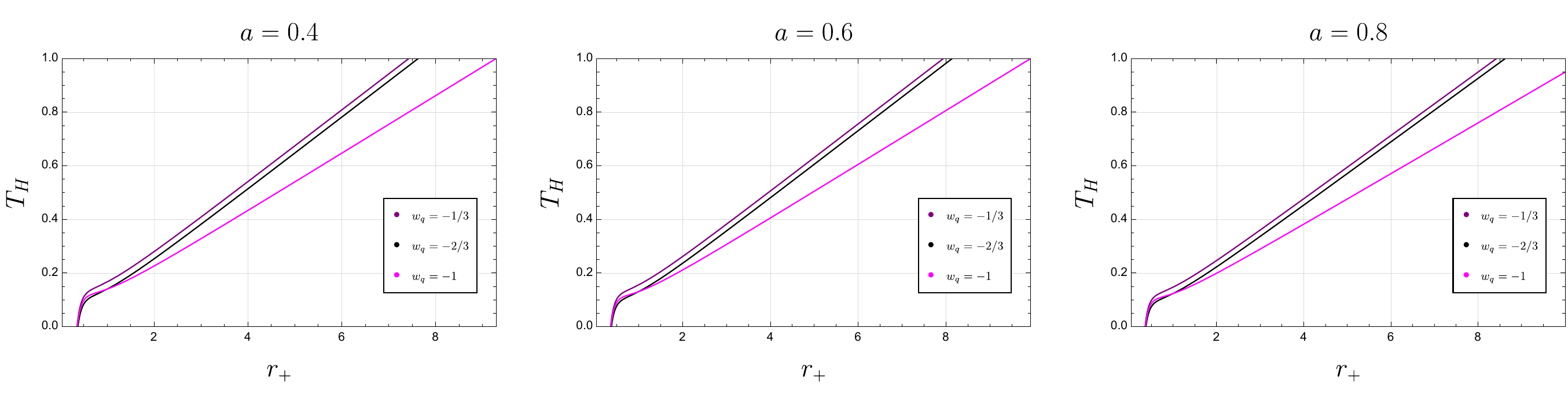}
      \caption{Plot of temperature versus event horizon radius $r_+$. With $m=0.4$, $Q=0.4$ and $N_q=-0.4$.}
        \label{f11}
\end{figure}

\begin{figure}[!htb]
     \centering
       \includegraphics[width=0.9\textwidth]{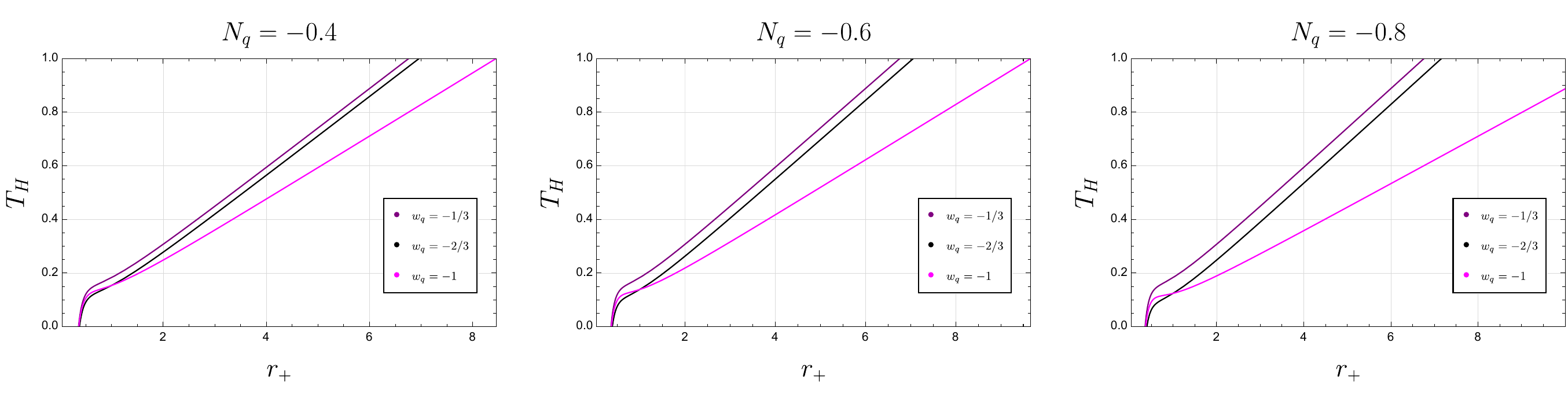}
      \caption{Plot of temperature versus event horizon radius $r_+$. With $m=0.4$, $Q=0.4$ and $a/l=0.4$.}
        \label{f12}
\end{figure}

The tunneling probability of a particle from inside to outside the black string horizon is given by \cite{Parikh:1999mf}
\begin{equation}
\Gamma\propto\frac{P_{\text{emission}}}{P_{\text{absorption }}}=\exp\left(-\frac{4}{\hslash}\text{Im}I\right)=\exp\left(-4\text{Im}W_{+}\right),
\end{equation}
which in terms of eq. (\ref{W+}) results in
\begin{equation}\label{ProbabilidadeTunelamento}
\Gamma=\exp\left[-\frac{4\pi\lambda\left(E-J_{1}\Omega_{+}+eA_{t}\left(r_{+}\right)\right)}{\left(\frac{2r_{+}}{l^{2}}+\frac{2ml}{r_{+}^{2}}-\frac{\left(3w_{q}+1\right)N_{q}}{r_{+}^{3w_{q}+2}}-\frac{2l^{2}Q^{2}}{r_{+}^{3}}\right)}\right].
\end{equation}
Comparing (\ref{ProbabilidadeTunelamento}) with $\Gamma = \exp[-\beta E]$ , where $\beta=(T_{H}^{-1})$, we identify the Hawking temperature
\begin{equation}\label{Hawking}
T_{H}=\frac{1}{4\pi\lambda}\left(\frac{2r_{+}}{l^{2}}+\frac{2ml}{r_{+}^{2}}-\frac{\left(3w_{q}+1\right)N_{q}}{r_{+}^{3w_{q}+2}}-\frac{2l^{2}Q^{2}}{r_{+}^{3}}\right),
\end{equation}
whose dependence is explicit in terms of the spacetime parameters and the radius of the event horizon, and by taking the proper limit, is in line with the results of the existing literature \cite{Ahmed:2011qh,Ali:2019mxs}. In Figures \ref{f7} and \ref{f8}, using three different values for $w_q$, we plot the Hawking temperature with respect to the variable $r_+$ and the parameters $m$ (Figure \ref{f7}) and $Q$ (Figure \ref{f8}) respectively. On the other hand, in Figures \ref{f7} and \ref{f9}, \ref{f10}, \ref{f11} and \ref{f12}, we consider the graphs of the Hawking temperature with respect to the variable $r_+$ using
three different values for $m$, $Q$ and $N_q$ respectively.In all of these figures, $N_{q}$ is given in units of $l^{3w_{q}+1}$ and $r_+$ in units of $l$.

\textcolor{black}{The analytical expression for $T_H$ indicates that its dominant contribution scales linearly with the horizon radius $r_+$, a feature that arises from the AdS contribution of the solution. Consequently, $T_H$ increases indefinitely with $r_+$, and there is no finite $r_+$ at which the temperature saturates. For $w_q = -1$, the quintessence correction term becomes proportional to $r_+$, thus introducing an additional linear modification. In contrast, for $w_q = -\frac{2}{3}$ the quintessence term reduces to a constant shift in $T_H$, and for $w_q = -\frac{1}{3}$ the correction vanishes entirely, so that the temperature profile is determined solely by the AdS, mass, and charge contributions. At small and intermediate $r_+$, the combined effects of the mass and charge terms with these distinct quintessence corrections lead to noticeably different temperature profiles. In summary, although $T_H$ increases continuously with $r_+$ in all cases, the influence of the quintessence field varies: it adds a linear correction for $w_q=-1$, a constant offset for $w_q=-\frac{2}{3}$, and no effect for $w_q=-\frac{1}{3}$.}

\section{Thermodynamics and stability}\label{secVI}
\textcolor{black}{The thermodynamic properties of the rotating charged black string surrounded by quintessence are influenced by the quintessence field and the rotation parameter. In this section, we analyze the entropy, temperature, and thermal stability of the black string, focusing on the effects of quintessence and charge.}

\textcolor{black}{The entropy of the black string is given by the Bekenstein-Hawking area law \cite{bekenstein1973black}, which states that the entropy is proportional to the area of the event horizon. For the rotating black string metric in (\ref{DiagMetric}), the horizon geometry at \( r = r_+ \) is described by:
\begin{equation}\label{SolHorizon_2D}
    ds_+^2 = \lambda^2 r_+^2 d\chi^2 + \frac{r_+^2}{l^2} dz^2.
\end{equation}}

\textcolor{black}{The proper area \( A \) of the horizon is computed by integrating over the angular coordinate and the string axis, this yields \( A = 2\pi \lambda r_+^2 / l \), which enables us to write the entropy per unit length along \( z \)
\begin{equation}\label{eq:entropy}
    S_+ = \frac{A}{4} = \frac{\pi \lambda r_+^2}{2l}.
\end{equation} This result matches \cite{Kumar:2023agi,Dehghani:2002rr,Bakhtiarizadeh:2023mhk}, confirming that the area law holds even in the presence of quintessence. Using the horizon condition \( f(r_+) = 0 \), the mass parameter \( m \) can be expressed as \begin{equation}\label{eq:parameter_m}
    ml = \frac{r_+}{2} \left( \frac{r_+^2}{l^2} + \frac{N_q}{r_+^{3w_q + 1}} + \frac{l^2 Q^2}{r_+^2} \right),
\end{equation} and substituting this result into the expression for the Hawking temperature obtained via tunneling of scalar particles (\ref{Hawking}) we obtain \begin{equation}
    T_{\text{H}} = \frac{3}{4\pi \lambda r_+} \left( \frac{r_+^2}{l^2} - \frac{w_q N_q}{r_+^{3w_q + 1}} - \frac{l^2 Q^2}{3r_+^2} \right).
\end{equation} This expression highlights the influence of the quintessence parameter \( N_q \), the equation of state parameter \( w_q \), and the charge \( Q \). For \( N_q = 0 \) and \( Q = 0 \), the temperature reduces to the Lemos black string result and when \( N_q \neq 0 \) and \( Q = 0 \), it agrees with the findings of \cite{Ali:2019mxs}. The charge \( Q \) introduces a term \(-\frac{l^2 Q^2}{3r_+^2}\), which decreases the temperature, reflecting the electrostatic effects, this result illustrates the role of charge in modifying the thermodynamic properties of the system.}

\textcolor{black}{We now examine the thermodynamic stability of the solution (\ref{SolII}), the black string is stable if its heat capacity is positive and unstable otherwise. It is given by
\begin{equation}
    C_{+} = \left( \frac{dm}{dT} \right)_{r=r_{+}},
\end{equation}
and using the expressions for \( m \) and \( T_{\text{H}} \),  in units of \( l \) we have
\begin{equation}
    C_+ = \frac{2\pi\lambda r_{+}^{2} \left( \frac{r_{+}^{2}}{l^{2}} - \frac{w_q N_q}{r_{+}^{3w_q + 1}} - \frac{l^2 Q^2}{3r_{+}^{2}} \right)}{\left( \frac{r_{+}^{2}}{l^2} + \frac{\left(3w_q + 2\right) N_q w_q}{r_{+}^{3w_q + 1}} + \frac{l^2 Q^2}{r_{+}^{2}} \right)}.
\end{equation}}

\textcolor{black}{For a given set of parameters, there exists a critical radius \( r_c \), where \( C \rightarrow \infty \). The equation for the critical radius is
\begin{equation}\label{critical_radii}
    r_{+}^{3w_q + 3} + l^{4} Q^2 r_{+}^{3w_q - 1} + \left(3w_q + 2\right) l^2 N_q w_q = 0,
\end{equation}
and this expression shows the contribution of the solution parameters (\ref{SolII}) to the stability. In the special case where \( Q = 0 \), we recover the result from \cite{Ali:2019mxs}, which allows us to obtain an explicit solution for \( r_c \)
\begin{equation}
    r_{+}^{c} = \left[ -\left( 3w_q + 2 \right) N_q w_q l^2 \right]^{\frac{1}{3w_q + 3}}.
\end{equation}}

\textcolor{black}{For \( Q \neq 0 \), we can compute the critical radius for different values of \( w_q = -1, -2/3, 1/3 \). Table \ref{tab:critical_radii} summarizes the critical radii $r_c^+$, as obtained from (\ref{critical_radii}), which represent the divergence points of the heat capacity $C_+$ and these divergences signal a change in the thermodynamic stability of the black string. Notably, for $w_q = -1$, real values of $r_c^+$ exist only under the condition \textcolor{black}{$N_{q}<-\frac{1}{l^{2}}$}, while for $w_q = -\frac{2}{3}$ and $w_q = -\frac{1}{3}$ the critical radii are complex, indicating that no physical phase transition occurs in those cases.}

\begin{figure}[!htb]
     \centering
       \includegraphics[width=0.9\textwidth]{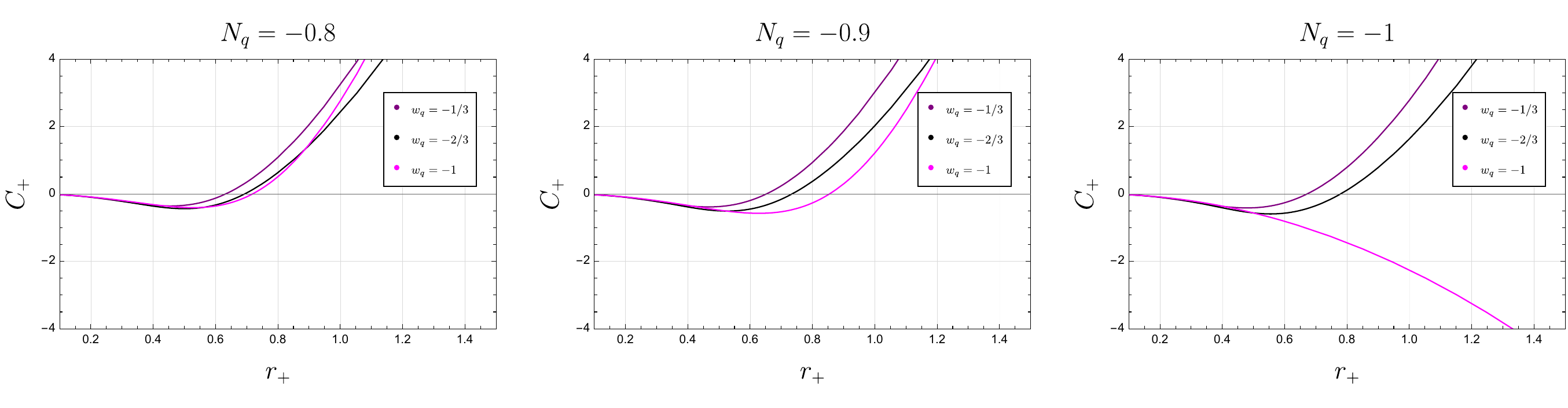}
      \caption{\textcolor{black}{Plots of the heat capacity \( C_+ \) vs. horizon radius \( r_+ \) for charged rotating black strings surrounded by quintessence, for different values of \( N_q \) in units of \( l^{3w_q + 1} \).
With $Q=0.4$ and $a/l=0.4$.}}
        \label{f13}
\end{figure}

\begin{figure}[!htb]
     \centering
       \includegraphics[width=0.9\textwidth]{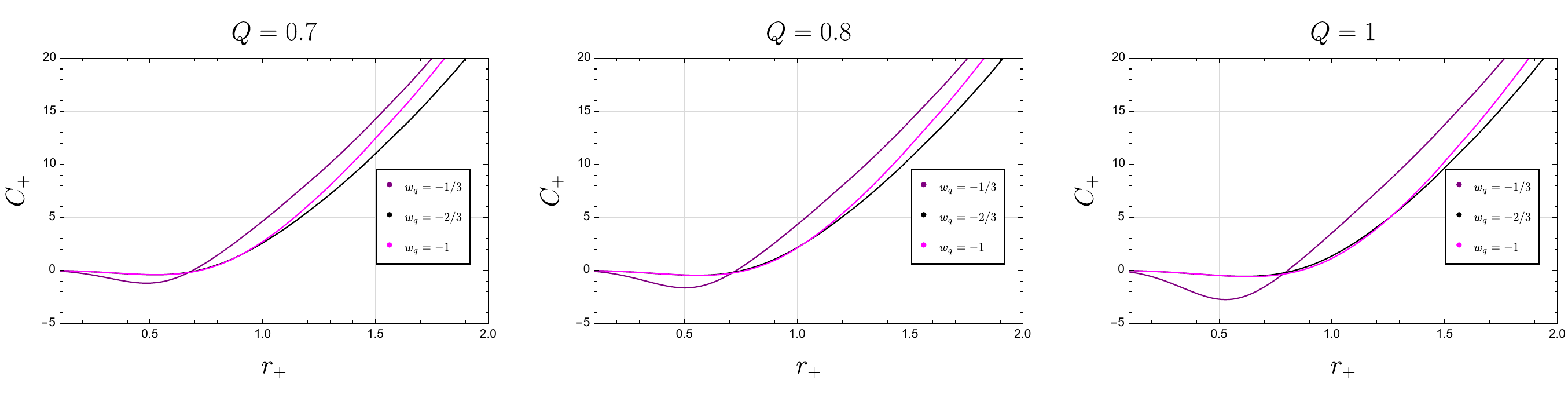}
      \caption{\textcolor{black}{Plots of the heat capacity \( C_+ \) vs. horizon radius \( r_+ \) for different values of \( Q \).
With $a/l=0.4$ and $N_q=-0.4$.}}
        \label{f14}
\end{figure}

\textcolor{black}{Figures \ref{f13} and \ref{f14} further illustrate the thermodynamic behavior by plotting $C_+$ versus the horizon radius $r_+$. In Figure \ref{f13} (with fixed $Q = 0.4$ and $a/l = 0.4$), variations in $N_q$ shift the location of the divergence in $C_+$, reflecting the influence of the quintessence field on the stability of the black string. Likewise, Figure \ref{f14} (with fixed $N_q = -0.4$ and $a/l = 0.4$) shows that changes in the charge $Q$ similarly affect the divergence point of $C_+$, thereby altering the stability regime. In both figures, the divergence of the heat capacity divides the parameter space into regions where $C_+$ is positive (indicating local thermodynamic stability) and regions where it is negative (indicating instability).}

\begin{table}[H]
\centering
\begin{tabular}{c|c|c}
\hline
$w_q$ &  $r_+^c$ &  Physical Validity \\ \hline
$-1$ & $\sqrt[4]{\frac{-l^4 Q^2}{l^2 N_q + 1}}$ & Real if \( N_q < -\frac{1}{l^2} \)  \\  
$-\frac{2}{3}$ & $\sqrt[4]{-l^{4} Q^{2}}$ & Complex (No physical solution) \\  
$-\frac{1}{3}$ & $\sqrt{\frac{1}{6}l^{2} N_q + \frac{1}{2} \sqrt{\frac{1}{9} l^4 N_q^2 - 4 l^4 Q^2}}$ & Complex (No physical solution) \\  
\hline
\end{tabular}
\caption{\textcolor{black}{Critical radii \( r_+^c \) as a function of \( N_q \), \( Q \), and \( l \) for different values of \( w_q \). Only the case \( w_q = -1 \) admits real solutions under the condition \( N_q < -\frac{1}{l^2} \). For \( w_q = -\frac{2}{3} \), since \( l > 0 \) and \( Q > 0 \), \( r_+^c \) becomes complex. For \( w_q = -\frac{1}{3} \), the conditions \( N_q^2 > 36 Q^2 \) and \( Q^2 < 0 \) must hold simultaneously, but since \( Q > 0 \) (given \( N_q < 0 \)), a real solution is not possible.}}
\label{tab:critical_radii}
\end{table}

\section{Conclusions}\label{secVII}
 We have investigated the use of an anisotropic stress--energy tensor associated with the Kiselev anisotropic fluid and electromagnetic field in Einstein field equations. 
Searching for solutions with cylindrical symmetry, we considered the metric of the black string in the form given by equation (\ref{MetricaCordaNegra}) . We obtained a solution for a charged black string immersed in an anisotropic fluid, which is asymptotically anti-de Sitter, and then the Hawking temperature. This solution encompasses the Lemos black string, which can be recovered in the absence of both fluid and charge.  We have assumed a negative cosmological constant in the form $\Lambda = - \frac{3}{l^2}$, as we can see in equation (\ref{HorizonteEventos}), the size of horizon is related to the charge, cosmological constant, parameter of equation of state and string rotation. These quantities provide a fourth-order algebraic mathematical expression for the size of the horizon when we choose particular values for the equation state parameter. 

\textcolor{black}{We analyzed the singularities associated with the solution obtained in this work, where different types of horizons can emerge depending on the values of the equation of state parameters. The calculation of the energy density associated with the Kiselev fluid indicated that the integration constant \( N_q \) is related to both the intensity of the energy density and the energy conditions of the solution. We explicitly determined the values of the curvature invariants and estimated the location of singularities for selected values of the parameter \( w_q \).}

Equation (\ref{HorizonteEventos}) gives us a general expression for the horizons of the black string, regarding the particular case of this equation associated with $w_q = -1$, several singularities occur at $l^2 N_q = -1$, so this value must be avoided in numerical expressions. As we show in Figure \ref{f5}, the event horizon disappears for $Q=0.8$ and $w = -1,-2/3,-1/3$. \textcolor{black}{Furthermore, we explored the conditions for extreme solutions of the event horizon, identifying cases where the inner and outer horizons coincide. These extreme solutions are shown to depend significantly on the mass, charge, and quintessence parameters. The analysis of the CTC radius revealed restrictions on the parameter space to prevent causality violations, while the rotation parameter was found to cause an angular deficit that modifies the spacetime geometry.}

\textcolor{black}{ We also investigated the conserved charges, such as mass, angular momentum, and electric charge, using the Brown-York formalism in Section \ref{secIV}. As it was expected, these charges reflect the contributions of the rotation parameter and the electromagnetic field.}

Hawking radiation spans a wide range of particles, including fermions and bosons. In particular, the emission of scalar fields has been investigated in various scenarios, such as spherically symmetric and cylindrical black holes. In this paper, we extend this analysis to the configuration of a rotating charged black string immersed in a Kiselev anisotropic fluid. Using the Hamilton-Jacobi method and complex integration, we solve the Klein-Gordon equation in the given background. Through the WKB approximation, we determine the tunneling probabilities through the event horizon. The temperature obtained is consistent with the existing literature taking the appropriate limits. The expression for the temperature depends on the rotation parameter $a$, which means that the Hawking temperature decreases/increases if $a$ increases/decreases, as we can see from equation (\ref{Hawking}). In addition, we show that the temperature increases for high values of $r_+$.    

\textcolor{black}{By performing a thermodynamic analysis, we calculate the entropy of the black string, which is in agreement with the literature. The heat capacity, as derived in the text, exhibits positive and negative values indicating stability of instability for certain values of the solution parameters.}

In future works, it can be fruitful to study additional thermodynamic properties such as the \textcolor{black}{black string}, pressure, and enthalpy. These quantities suggest the analysis of the properties of the extended phase space that can be derived from the spacetime metric obtained in this work. Regarding the singularity present in this black hole, we can study a bounce mechanism to regularize this solution similar to the approach used in \cite{franzin2021charged}.

\section{Data availability statement}
No new data were created or analysed in this study.

\section{Acknowledgements}
L.G.B would like to thank CAPES (Process number: 88887.642857/2021-00 and 88887.96\\8290/2024-00) for the financial support. L.C.N.S would like to thank FAPESC for financial support under grant 735/2024. V.H.M.R. would like to thank CAPES (Process number: 88887.483316/2020-00 and 88887.816474/2023-00) for the financial support. C. C. B. Jr. would like to thank CNPq (Conselho Nacional de Desenvolvimento Cient\'ifico e Tecnol\'ogico).

\bibliographystyle{ieeetr}
\bibliography{sample}

\appendix
\section{Riemann Tensor Components}\label{Riemann_Tensor_Components}

\textcolor{black}{In this appendix, we present the components of the Riemann tensor associated with the general metric used in this work:
\begin{equation}
    ds^{2}=-f(r)dt^{2}+\frac{dr^{2}}{f(r)}+r^{2}d\phi^{2}+\frac{r^{2}}{l^{2}}dz^{2}.
\end{equation}}
\textcolor{black}{These components were employed in the calculations leading to the Kretschmann scalar discussed in the main text.}
\textcolor{black}{For this metric, they are given by
\begin{align}
    R_{trtr} &= R_{rtrt} = -R_{trrt} = -R_{rttr} = \frac{1}{2} f''(r), \\
    R_{t\phi t\phi} &= R_{\phi t\phi t} = -R_{t\phi\phi t} = -R_{\phi tt\phi} = \frac{1}{2} r f(r) f'(r), \\
    R_{tztz} &= R_{ztzt} = -R_{tzzt} = -R_{zttz} = \frac{1}{2} r l^{-2} f(r) f'(r), \\
    R_{r\phi r\phi} &= R_{\phi r\phi r} = -R_{r\phi\phi r} = -R_{\phi rr\phi} = -\frac{1}{2} r \frac{f'(r)}{f(r)}, \\
    R_{rzrz} &= R_{zrzr} = -R_{rzzr} = -R_{zrrz} = -\frac{1}{2} r l^{-2} \frac{f'(r)}{f(r)}, \\
    R_{\phi z\phi z} &= R_{z\phi z\phi} = -R_{\phi zz\phi} = -R_{z\phi\phi z} = -r^{2} l^{-2} f(r),
\end{align}}
\textcolor{black}{and as it is well known, they characterize the curvature of the spacetime. By inspecting these expressions it is possible to observe the presence of singularities. }

\section{Metric Transformation}\label{Metric_Transformation}
\textcolor{black}{In this appendix, we derive the transformation of the metric under the coordinate change \( \phi = \chi + \Omega t \). Considering a general metric of the form \cite{Natario:2008ej,Sui:2023rfh,Israel:1970kp,Frolov:2014dta}
\begin{equation}
    ds^{2} = g_{tt} dt^{2} + g_{rr} dr^{2} + 2 g_{t\phi} dt d\phi + g_{\phi\phi} d\phi^{2} + g_{zz} dz^{2},
\end{equation}
then introducing the differential transformation \( d\phi = d\chi + \Omega dt \) and expanding the terms, we obtain
\begin{equation}
    ds^{2} = (g_{tt} + 2\Omega g_{t\phi} + g_{\phi\phi}\Omega^{2}) dt^{2} + g_{rr} dr^{2} + 2 (g_{t\phi} + \Omega g_{\phi\phi}) dt d\chi + g_{\phi\phi} d\chi^{2} + g_{zz} dz^{2}.
\end{equation}}

\textcolor{black}{By imposing the condition that the off-diagonal term vanishes, we find
\begin{equation}
    \Omega = -\frac{g_{t\phi}}{g_{\phi\phi}},
\end{equation}}

\textcolor{black}{and substituting this result, the metric simplifies to
\begin{equation}
    ds^{2} = \left(g_{tt} - \frac{g_{t\phi}^{2}}{g_{\phi\phi}}\right) dt^{2} + g_{rr} dr^{2} + g_{\phi\phi} d\chi^{2} + g_{zz} dz^{2}.
\end{equation}}

\textcolor{black}{For the charged black string solution immersed in an anisotropic quintessence fluid, as obtained in Eq. (\ref{SolII}),
\begin{equation}
\begin{split}
ds^{2} &= -\left(f(r)\lambda^{2} - r^{2} \frac{a^{2}}{l^{4}}\right) dt^{2} + \frac{dr^{2}}{f(r)} + \left(r^{2} \lambda^{2} - f(r) a^{2}\right) d\phi^{2} \\
&\quad  + \left(f(r) \lambda a - r^{2} \lambda \frac{a}{l^{2}}\right) 2 dt d\phi + \frac{r^{2}}{l^{2}} dz^{2},
\end{split}
\end{equation}
where
\begin{equation}
    f(r) = \frac{r^{2}}{l^{2}} - \frac{2ml}{r} + \frac{N_{q}}{r^{3w_{q}+1}} + \frac{l^{2} Q^{2}}{r^{2}},
\end{equation}
we identify the metric components as
\begin{align}
    g_{tt} &= -\left(f(r)\lambda^{2} - r^{2} \frac{a^{2}}{l^{4}}\right),\quad g_{rr} = \frac{1}{f(r)}, \\
    g_{\phi\phi} &= \left(r^{2} \lambda^{2} - f(r) a^{2}\right), \quad g_{t\phi} = \left(f(r) \lambda a - r^{2} \lambda \frac{a}{l^{2}}\right), \quad g_{zz} = \frac{r^{2}}{l^{2}}.
\end{align}}

\textcolor{black}{After performing the coordinate transformation, the metric can be rewritten in a diagonal form:
\begin{equation}\label{DiagMetric}
    ds^{2} = -F(r) dt^{2} + \frac{dr^{2}}{G(r)} + H(r) d\chi^{2} + K(r) dz^{2},
\end{equation}
where the functions are defined as
\begin{align}
    F(r) &= \frac{f(r) r^{2}}{r^{2} \lambda^{2} - f(r) a^{2}},\quad H(r) = r^{2} \lambda^{2} - f(r) a^{2}, \\
    G(r) &= f(r), \quad K(r) = \frac{r^{2}}{l^{2}}.
\end{align}}

\textcolor{black}{The diagonal form of the metric given in (\ref{DiagMetric}) is particularly useful for calculations involving field dynamics in this spacetime, angular deficit calculations, and thermodynamic quantities such as the entropy.}

\section{Relativistic Hamilton-Jacobi Equation via the WKB Method}\label{Relativistic_Hamilton_Jacobi_Equation_via_the_WKB_Method}

\textcolor{black}{In this section, we derive the relativistic Hamilton--Jacobi equation using the Wentzel-Kramers-Brillouin (WKB) approximation \cite{Parikh:1999mf,Gomes:2018oyd}. We begin by considering the Klein-Gordon equation in a curved spacetime background with an electromagnetic field
\begin{equation}
    \frac{1}{\sqrt{-g}}\left(\partial_{\alpha}-\frac{ie}{\hslash}A_{\alpha}\right)\left[\sqrt{-g}g^{\alpha\beta}\left(\partial_{\beta}-\frac{ie}{\hslash}A_{\beta}\right)\Psi\right]-\frac{m^{2}}{\hslash^{2}}\Psi=0
\end{equation}
where $m$ is the mass of the scalar particle, and $e$ is its charge. Expanding the terms inside the brackets, we obtain:
\begin{multline}
 g^{\alpha\beta}\partial_{\alpha}\partial_{\beta}\Psi-
\frac{2ie}{\hslash}g^{\alpha\beta}A_{\beta}\partial_{\alpha}\Psi
-\frac{e^{2}}{\hslash^{2}}g^{\alpha\beta}A_{\alpha}A_{\beta}\Psi
-\frac{m^{2}}{\hslash^{2}}\Psi \\
+\frac{1}{\sqrt{-g}}\partial_{\alpha}\left(\sqrt{-g}g^{\alpha\beta}\right)\partial_{\beta}\Psi
-\frac{ie}{\hslash}\frac{1}{\sqrt{-g}}\partial_{\alpha}\left(\sqrt{-g}g^{\alpha\beta}\right)A_{\beta}\Psi
-\frac{ie}{\hslash}g^{\alpha\beta}\partial_{\alpha}A_{\beta}\Psi=0.
\end{multline}}

\textcolor{black}{Applying the WKB ansatz
\begin{equation}
\Psi\left(t,r,\phi,z\right)=\exp\left[\frac{i}{\hslash}I\left(t,r,\phi,z\right)\right],
\end{equation}
we substitute into the equation, leading to the expression
\begin{multline}
-\frac{1}{\hslash^{2}}g^{\alpha\beta}\partial_{\alpha}I\partial_{\beta}I
+\frac{2e}{\hslash^{2}}g^{\alpha\beta}A_{\beta}\partial_{\alpha}I
-\frac{e^{2}}{\hslash^{2}}g^{\alpha\beta}A_{\alpha}A_{\beta}-\frac{m^{2}}{\hslash^{2}} \\
+\frac{i}{\hslash}\frac{1}{\sqrt{-g}}\partial_{\alpha}\left(\sqrt{-g}g^{\alpha\beta}\right)\partial_{\beta}I
-\frac{ie}{\hslash}\frac{1}{\sqrt{-g}}\partial_{\alpha}\left(\sqrt{-g}g^{\alpha\beta}\right)A_{\beta}
-\frac{ie}{\hslash}g^{\alpha\beta}\partial_{\alpha}A_{\beta}=0.
\end{multline}}

\textcolor{black}{Multiplying by $\hslash^{2}$ and taking the semiclassical limit $\hslash\to0$, the imaginary terms vanish, yielding the relativistic Hamilton--Jacobi equation:
\begin{equation}
    g^{\alpha\beta}\left(\partial_{\alpha}I\partial_{\beta}I-2eA_{\beta}\partial_{\alpha}I+e^{2}A_{\alpha}A_{\beta}\right)+m^{2}=0.
\end{equation}
This equation governs the motion of relativistic scalar particles in curved spacetime under the influence of an electromagnetic field, serving as the foundation for the semiclassical analysis of quantum tunneling and related phenomena.}

\end{document}